\documentclass[twocolumn]{aa}
\usepackage{graphicx}
\usepackage[colorlinks=true, linkcolor=blue, citecolor=blue, urlcolor=gray]{hyperref}
\usepackage{natbib}
\usepackage{orcidlink}
\usepackage[varg]{txfonts}
\usepackage{upgreek}
\usepackage{xcolor}
\usepackage{xspace}

\newcommand{\mum}{\ensuremath{\upmu}m\xspace}
\newcommand{\csixty}{C$_{60}$\xspace}

\begin{document}

\title{JWST observations of photodissociation regions. \\
IV. Carbonaceous emission band sub-components in NGC 7023 have distinct spatial distributions}

\author{D. Van De Putte\inst{\ref{uwo}, \ref{stsci} \orcidlink{0000-0002-5895-8268}} \and
        K.~D. Gordon \inst{\ref{stsci}, \ref{gent} \orcidlink{0000-0001-5340-6774}}    \and
        K. Misselt \inst{\ref{az}}        \and
        A.~N. Witt \inst{\ref{toledo} \orcidlink{0000-0003-0760-4483}}       \and
        A. Abergel \inst{\ref{ias} \orcidlink{0000-0003-2364-2260}}        \and
        A. Noriega-Crespo \inst{\ref{stsci} \orcidlink{0000-0002-6296-8960}} \and
        P. Guillard \inst{\ref{sorbonne} \orcidlink{0000-0002-2421-1350}}       \and
        M. Zannese \inst{\ref{ias} \orcidlink{0009-0002-6069-0907}}        \and
        M. Elyajouri \inst{\ref{stsci} \orcidlink{0000-0002-6086-2337}}  \and
        B. Trahin \inst{\ref{stsci}, \ref{ias} \orcidlink{0000-0001-5875-5340}}      \and
        P. Dell'ova \inst{\ref{ias}}       \and
        M. Baes \inst{\ref{gent} \orcidlink{0000-0002-3930-2757}} \and
        P. Klaassen \inst{\ref{UKATC} \orcidlink{0000-0001-9443-0463}}
        }

\institute{
Department of Physics \& Astronomy, The University of Western Ontario, London ON N6A 3K7 Canada\label{uwo} \and
Space Telescope Science Institute, 3700 San Martin Drive, Baltimore, MD, 21218, USA \label{stsci} \and
Sterrenkundig Observatorium, Universiteit Gent, Krijgslaan 281 S9, B-9000 Gent, Belgium \label{gent} \and
Steward Observatory, University of Arizona, Tucson, AZ 85721-0065, USA \label{az} \and
Ritter Astrophysical Research Center, University of Toledo, Toledo, OH 43606, USA \label{toledo} \and
Institut d'Astrophysique Spatiale, Université Paris-Saclay, CNRS, 91405 Orsay, France \label{ias} \and
Sorbonne Université, CNRS, Institut d'Astrophysique de Paris, 98\,bis bd Arago, 75014 Paris, France\label{sorbonne} \and
United Kingdom Astronomy Technology Centre, Edinburgh, GB, UK\label{UKATC}
}

\abstract
{
    The northwest photodissociation region (PDR) of the NGC\,7023 reflection nebula was observed with JWST spectroscopy, revealing the carbonaceous emission bands in high detail.
    Compared to other PDRs (Horsehead and Orion Bar) the softer radiation field driving NGC\,7023 results in an extended atomic hydrogen region and more pronounced spatial variations of the band profiles.
    Its weaker thermal dust continuum makes NGC\,7023 an ideal target to study the 16-18~\mum emission and its relation to the main emission bands at 3.3, 3.4, 5.2, 5.7, 6.2, 7.7, 8.6, 11.3, and 12.7~\mum.
}
    {
    We perform a spatially resolved study to reveal which emission bands originate from the same or co-spatial populations of small carbonaceous emitters.
    }
    {
    We apply a spectral decomposition with PAHFIT to around 500 extracted spectra, and produce maps of the individual sub-components of the carbonaceous emission bands.
    The spatial resolution of around 0.7~mpc spatially resolves the dissociation front.
    }
    {
    Nearly all feature maps peak at the dissociation front (DF1), while the emission in the atomic PDR region (ATM) varies strongly between the bands and their sub-components.
    We organize the spatial distributions into three categories based on the intensity ratio in ATM relative to DF1.
    Most bands are of type I (low ATM/DF1; 3.3, 3.4, 5.2, 5.7, 11.3~\mum) or II (medium ATM/DF1; 16.2, 7.7, 8.6, 12.7, 16.4~\mum), while only few are of type III (high ATM/DF1; 11.0, 17.4~\mum).
    The decompositions and maps of the 5.7, 7.7, 11.3 and 12.7~\mum bands reveal that their spectral profiles consist of blue-side sub-components with a type III spatial distribution, while redder sub-components are of type I or II.
    The 17.4~\mum band correlates strongly with these blue sub-components and the 8.6 and 11.0~\mum features, which trace charged carriers.
    }
    {
    The different 17.4 and 16.4~\mum spatial distributions indicate at least two different populations contributing to the 16-18~\mum range.
    The strongly differing spatial distribution types of carbonaceous band sub-components reveal a connection between the 11.0 and 17.4~\mum bands and the spectral profiles of the 5.7, 7.7, 11.3, and 12.7~\mum bands.
    The maps indicate that these profiles will continue evolving as the central cavity is approached.
    Since C$_{60}$ emission was previously detected in the cavity, we speculate that the population of emission carriers could be in an intermediate photochemical evolution stage of fullerene formation.
    }


\titlerunning{Carbonaceous emission band sub-components in NGC\,7023 have distinct spatial distributions}

\maketitle
\nolinenumbers

\section{Introduction}

Photodissociation regions (PDRs) are the transition between the ionized and molecular regions in the interstellar medium (ISM), and appear at the edges of interstellar clouds when they are irradiated by a far-ultraviolet (FUV) source.
Typically there is a sharp H{\sc{ii}}-to-H{\sc{i}} transition or ionization front (IF), and an H{\sc{i}}-to-H$_2$ transition or dissociation front (DF), of which the location and length scale are driven by the attenuation of the FUV radiation field and the heating, cooling, and photochemical processes it influences \citep{Hollenbach1997, Hollenbach1999, Wolfire2022}.
In nearby PDRs (distance $\lesssim 1000$ pc) with an edge-on orientation, these transitions can be spatially resolved, which is why they are considered ideal laboratories to study interstellar gas and dust and its response to changes in the environment (radiation field, density, temperature, molecular fraction).

PDRs are typically very bright in the near-infrared (NIR) and mid-infrared (MIR), as the UV excitation of small carbonaceous species in the ISM drives radiative cascades through their vibrational states, resulting in a group of emission bands spanning from around 3 to 20~\mum \citep{Leger1984, Allamandola1985}.
Most of these bands are associated with vibrational modes of bonds found in polycyclic aromatic hydrocarbons (PAHs), and they are also referred to as the aromatic infrared bands (AIBs).
The brightest emission features of this nature are found at 3.3, 5.2, 6.2, 7.7, 8.6, 11.0, 11.3, 12.7, and 16.4 \mum, and each of these bright bands appears to exhibit shape variations, or consists of multiple components \citep{Peeters2002,Peeters2004,vanDiedenhoven2004,Chown2024, VanDePutte2024}.
JWST revealed one of the most detailed views of the AIB spectrum with its observations of the Orion Bar PDR, revealing numerous weaker sub-components that blend with the main bands and affect the shape of their emission profiles \citep{Chown2024}.
Diagnostics based on the intensities of these bands are used in observational studies to detect the evolution of their physical properties such as the size or charge of the carriers, and these changes are commonly interpreted as tracers for the environment, such as radiation field or density \citep[e.g.,][]{Galliano2008,Pilleri2012,Stock2017}.
Highly spatially resolved studies of edge-on PDRs minimize the blending between such different conditions within the beam size or along the line of sight.

NGC\,7023 is an isolated reflection nebula for which HD200755, the B2Ve binary star at its center, is the single driver of the illumination and the PDR physics.
It is of particular interest for the study of fullerenes (C$_{60}$) in space, observed in the central cavity \citep{Sellgren2007, Sellgren2010, Berne2012}.
Several PDRs and their DFs are observed as bright illuminated ridges bordering this cavity.
Integral field spectroscopy mosaics were observed with NIRSpec and MIRI as part of the JWST Guaranteed Time Observations (GTO) program \#1192 \citep{Misselt2025}.
In the context of the available JWST observations of PDRs, NGC\,7023 is driven by a soft radiation field of moderate intensity ($T_\text{eff} \sim 18000$ K, $G_0 \sim 2.6 \times 10^3$; \citealt{Chokshi1988}).
Comparatively, the radiation field driving the Horsehead is hard with a low intensity ($T_\text{eff} \sim 33000$ K, $G_0 \sim 100$; \citealt{Habart2005}), while that driving the Orion Bar is hard with a high intensity ($T_\text{eff} \sim 39000$ K, $G_0 \sim 2 \times 10^4$; \citealt{Marconi1998, Peeters2024}),
Here, $G_0$ is the radiation field in units by \citet{Habing1968}, and $G_0 = 1.7$ represents the average radiation field in the local ISM \citep{Draine1978}.

The above radiation field properties result in an extended atomic hydrogen region of the PDR, while an IF and the typical H recombination lines are not observed in our field of view.
The emission bands are much brighter compared to cases such as the Horsehead \citep{Misselt2025}, while the thermal dust continuum longward of 15~\mum is much weaker than that observed in the Orion Bar \citep{VanDePutte2024}.
The above properties allow for a highly detailed decomposition of the AIBs over the entire 3-20 \mum wavelength range for individual spatial elements of the integral field units (IFUs), and the limited dust continuum emission makes NGC\,7023 especially suitable to study the 16-18~\mum complex.
The general importance of the 16-18~\mum complex lies in their connection to C-C-C bending modes, as discussed in previous Spitzer-based works \citep{Peeters2004, Boersma2010}.
Some studies in NGC\,7023 revealed that the 16.4 and 17.4~\mum spatial distributions may be different, possibly related to the presence of different ionization stages of PAHs \citep{Shannon2015, Shannon2016}.
The spectro-spatial data and emission feature maps from the Spitzer era \citep{Werner2004, Sellgren2010} also showed that C$_{60}$ is present closer to the star, based on the spatial distribution of its main emission feature at 18.9~\mum, which peaks in the central cavity.
Another characteristic C$_{60}$ feature at 17.4~\mum is blended with the AIB at the same wavelength, and as such exhibits two local maxima: one near the peak of the 18.9~\mum emission, and one near the bright northwest (NW) filaments (the DFs of the PDR).
The region where C$_{60}$ is observed does not fall within the field of view of the JWST spectroscopy present here, though intermediate species may be observed between the C$_{60}$ region and the DF \citep{Berne2012}.
However, at the spatial resolution of Spitzer, the separation between the emission (sub-)components is not clearly established.

We aim to leverage the JWST spectroscopy to investigate the spatial distributions of the carbonaceous emission features in detail.
We place a particular focus on the 16.4 and 17.4~\mum bands, and make use of the entire 3 to 20 micron wavelength range to decompose the other prominent feature complexes, and show which bands or components have spatial distributions that match these 16-18~\mum emission sub-components.
In Sect.~\ref{sec:data}, we summarize our processing of the JWST NIRSpec and MIRI IFU data.
The methods of Sect.~\ref{sec:analysis} decompose the spectra of each spaxel using the PAHFIT tool, and the results are presented as maps of each sub-component in Sect.~\ref{sec:results}.
In Sect.~\ref{sec:discussion} we discuss which spatial distributions are most similar and may therefore originate from the same populations.
We compare these findings to previous work to provide a number of possible interpretations.

\section{Data}
\label{sec:data}

\begin{figure*}[tb]
    \centering
    \includegraphics[width=0.45\linewidth]{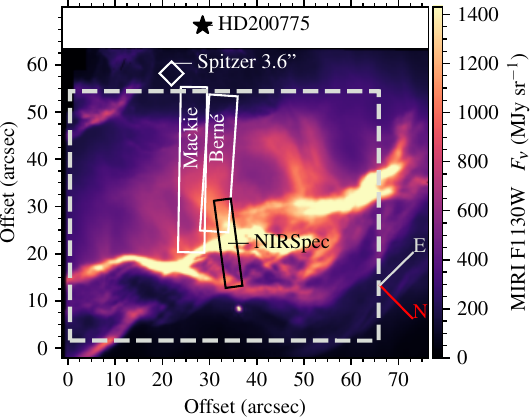}
    \hspace{.5cm}
    \includegraphics[width=0.48
\linewidth]{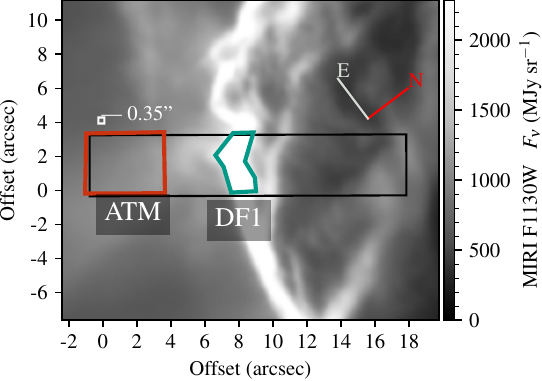}
    \caption{Left panel: Position of HD200775 relative to the NGC\,7023 NW filament and footprints of the NIRSpec observations.
    The background is the F1130W map from the MIRI imager observations of the same program \citep{Misselt2025}.
    The dashed rectangle indicates the coverage of Spitzer-based spectroscopy and feature maps \citep{Werner2004, Shannon2015}, which includes coverage closer to the star, where C$_{60}$ was observed.
    The labeled footprints are the regions of interest for the studies by \citet{Berne2012} and \citet{Mackie2015}.
    The diamond (a 3.6" square)  indicates the typical spectral extraction size for Spitzer-based studies.
    Right panel: Zoom-in of the same image, showing the two extraction apertures of \citet{Misselt2025} used for the demonstration in Figs.~\ref{fig:specandfit} and \ref{fig:specandfit2}, and the size of a single 0.35\arcsec~spaxel of our spectral extractions (Sect.~\ref{sec:data}).
    The edge of the mosaic is around 37\arcsec~away from the star.
    }
    \label{fig:footprint}
\end{figure*}

We make use of the NGC\,7023 data from guaranteed time observation program \#1192 (GTO-1192) for JWST \citep{Misselt2025}.
The observing strategy is similar to that of the PDRs4All program for the Orion Bar \citep{Berne2022}, where NIRSpec and MIRI IFU mosaics spatially resolve the variations of the spectral features.
The footprint of the NIRSpec mosaic is illustrated in Fig.~\ref{fig:footprint}, with a MIRI F1130W image of the same program displayed in the background.
The additional rectangles shown in Fig.~\ref{fig:footprint} are two regions studied previously using Spitzer data \citep{Berne2012, Mackie2015}; in those studies, the 18.9~\mum emission of C$_{60}$ was detected on the star-facing side of the respective rectangles.
The field of view of the JWST spectroscopy does not extend as far toward the central star, and therefore no detection of the 18.9~\mum feature is expected.

A detailed overview of the spectroscopic observing strategy, the data reduction, and a spectral inventory are provided by \citet{Misselt2025}.
In summary, they employed a few extra steps including an improved flagging of outliers for a cleaner end result, and a world coordinate system (WCS) adjustment based on the imaging data.
To apply our all-wavelength, per-spaxel spectral decomposition (Sect.~\ref{sec:pahfitcube}), we need data cubes for which the spatial grids are identical.
We built a total of 13 separate data cubes (NIRSpec IFU G395M and MIRI MRS Channel 1 through 4, SHORT MEDIUM and LONG bands), by starting from the CRF files produced by \citet{Misselt2025}, and running the cube building step of the JWST pipeline with custom arguments.
The NIRSpec IFU cubes have the smallest spatial extent, and MIRI MRS Channel 1 covers a similar area.
To ensure full wavelength coverage and to avoid oversampling the longer wavelength channels, each cube was built to the resolution of the MIRI MRS channel 4 (the largest spaxel size, $0\farcs35$) while the analysis is limited to the NIRSpec IFU footprint ($3\farcs3 \times 21\arcsec$ or $9\times52$ spaxels).
This combination of spaxel size and footprint dimensions results in around 500 spectral extractions for our analysis.
The footprint and the size of the spaxel apertures are illustrated in the right panel of Fig.~\ref{fig:footprint}.

To correct the astrometry based on NIRCam data, \citet{Misselt2025} applied a shift of around 6 spaxels to the NIRSpec data, parallel to the long axis of the mosaic and toward the star.
At the NIRSpec data cube resolution ($0\farcs1$), this shift corresponds to around $0\farcs6$, or about 2 spaxels in our lower-resolution ($0\farcs35$) cube.
To apply the same correction, we modified the WCS of each CRF file prior to applying the cube build step.
The shift for the MIRI IFU cubes was minimal.
In practice, this WCS correction reduces the gap between the emission peaks at the DF when the NIRSpec and MIRI IFU maps are compared.

We also extracted spectra for two larger apertures: the ``ATM'' (atomic PDR) and ``DF1'' (dissociation front 1) regions defined by \citet{Misselt2025} and shown in the right panel of Fig.~\ref{fig:footprint}.
We use these spectra with high signal-to-noise ratio to illustrate the spectral variations in the top panels of Figs.~\ref{fig:specandfit} and \ref{fig:specandfit2}, and to demonstrate the decomposition method explained in the next section.
To apply the decomposition over the complete wavelength range of interest (3.2-26~\mum), a single data cube covering the combined NIRSpec and MIRI wavelength range is needed.
The jumps in flux between the channels are small (a few \%), and to combine the cubes that are already on a common spatial grid by construction (as described above), we used an algorithm that smoothly blends the spectral segments by using a sliding weighted mean in the overlap regions \citep{VanDePutte2024}.

\section{Method}
\label{sec:analysis}

\subsection{Spectral decomposition with PAHFIT}
\label{sec:decompositionmethod}

To decompose the spectra, we make use of the latest version of the PAHFIT tool \citep{Smith2007}.
While originally written in IDL (interactive data language), PAHFIT was recently redeveloped in Python to enhance its usability for the analysis of JWST data.\footnote{\url{https://github.com/PAHFIT/pahfit/}}
For this work specifically, we use the same version of PAHFIT as presented by \citet{VanDePutte2025}.
A crucial part of applying PAHFIT to a given spectrum is the use of a suitable ``science pack'', which is the list of components to be fit and their parameters.
The PAHFIT model represents the AIBs with a combination of Drude profiles for which the amplitudes are fit independently, with the central wavelengths and the full width at half maximum (FWHM) listed in the science pack.
To set up a suitable AIB configuration, we start with the parameters provided by the ``PDR pack'' presented by \citet{VanDePutte2025}, which was set up based on JWST spectra of the Orion Bar \citep{Peeters2024, Chown2024, VanDePutte2024}.
For a detailed account of how the PDR science pack parameters were chosen, we refer to the appendix of \citet{VanDePutte2025}.
Since the signal-to-noise and AIB-to-continuum ratios of the 16-18~\mum complex are much higher in NGC\,7023 compared to the Orion Bar, a more detailed view of these emission bands is revealed.
We revised this part of the PDR pack to obtain reliable fits of the 17.4~\mum feature (see Sect.~\ref{sec:decompositionresults}).
The changes will be included in an updated version of the PDR pack and hereafter, each mention of the PDR pack refers to the one updated in this work.

The PDR pack is supported by additional PAHFIT components that represent the thermal continuum of the dust.
The standard PAHFIT approach uses modified blackbody (MBB) curves for which the intensity scales as $\lambda^{-2} B_\nu(\lambda, T)$, where $B_\nu$ is the Planck function evaluated in $F_\nu$ units (e.g., MJy / sr) at a wavelength $\lambda$ for a certain temperature $T$ \citep{Smith2007}.
\citet{VanDePutte2025} introduced an alternate continuum shape which includes the silicate emission around 10 and 20~\mum, to model the steep 15-26~\mum continuum of the Orion Bar.
For NGC\,7023, we find that using a set of standard PAHFIT MBB of temperatures $T_i$ (intensity $\sim \lambda^{-2} B_\nu(\lambda, T_i)$) produces an acceptable continuum model that does not suffer of the mismatch near 15~\mum, as the continuum of NGC\,7023 is not as steeply increasing toward longer wavelengths.
We include continuum components with the following values for $T_i$:
    35,
     40,
     45,
     50,
     58,
     65,
     78,
     90,
     113,
     135,
     168,
     200,
     250,
     300,
     and
     400 K.
A stellar continuum component was not included.
We note that the PAHFIT attenuation model was not included in the fit, even though the molecular region is likely affected by dust extinction, as would be expected given the presence of H$_2$O and CO$_2$ ice absorption \citep{Misselt2025}.
As noted in previous work \citep{Peeters2017, VanDePutte2025}, the inclusion of attenuation can lead to degeneracies which cause large jumps in the attenuation between different spatial elements.

Analogous to \citet{VanDePutte2025}, we removed the emission lines from the data, rather than including them in the model and fitting all of them with PAHFIT, as this can improve the convergence and consistency.
This was done by masking out small parts of the wavelength range, typically within two times the FWHM of each line.

\begin{figure*}[tbh]
    \centering
    \includegraphics[scale=0.9]{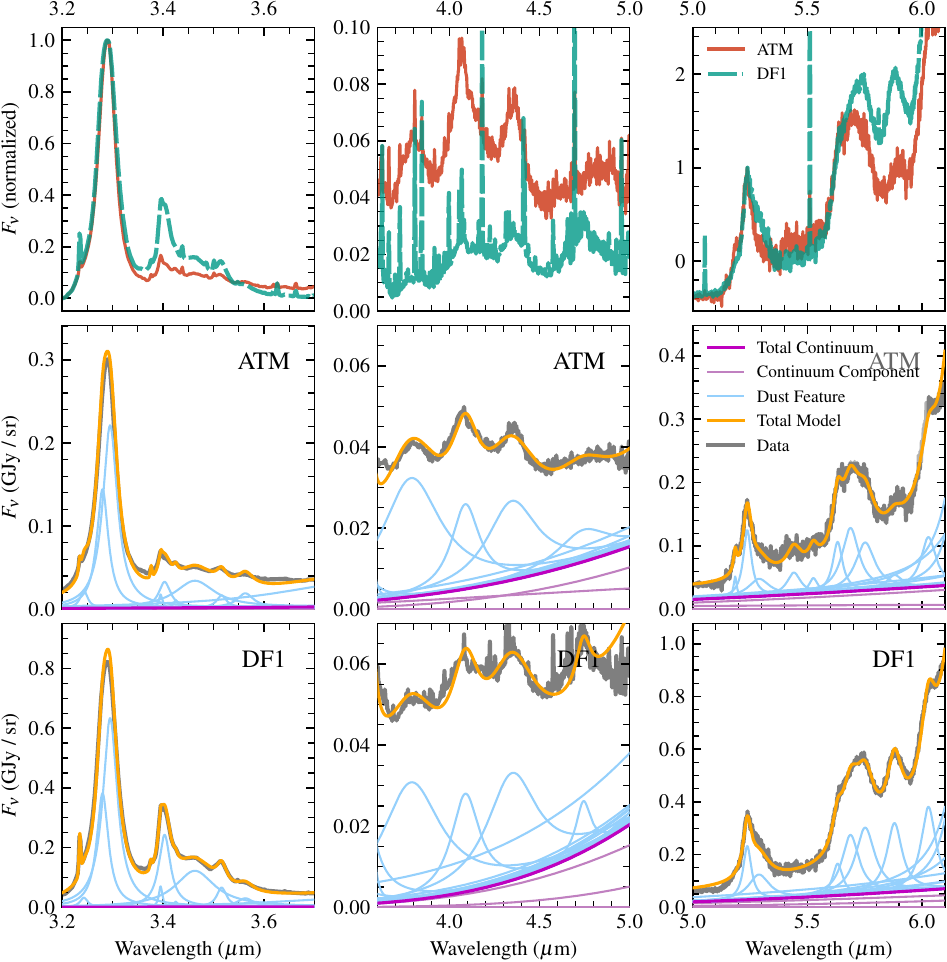}
    \caption{Zoomed-in view of the emission complexes in the ATM and DF1 regions of NGC\,7023, and the corresponding PAHFIT decompositions.
    The aperture-extracted spectra (defined in Sect.~\ref{sec:data} and Fig.~\ref{fig:footprint}) are shown in the top panels with a normalization and offset that match the base and peak of a certain feature (first and second columns scaled to 3.3~\mum, third column to 5.2~\mum); these still have the emission lines included.
    The gray lines in the bottom two rows are the modified ATM and DF1 spectra (emission lines removed), used as input for PAHFIT.
    The components of the decomposition are shown as labeled in the legend.}
    \label{fig:specandfit}
\end{figure*}

\begin{figure*}
    \centering
    \includegraphics[scale=.9]{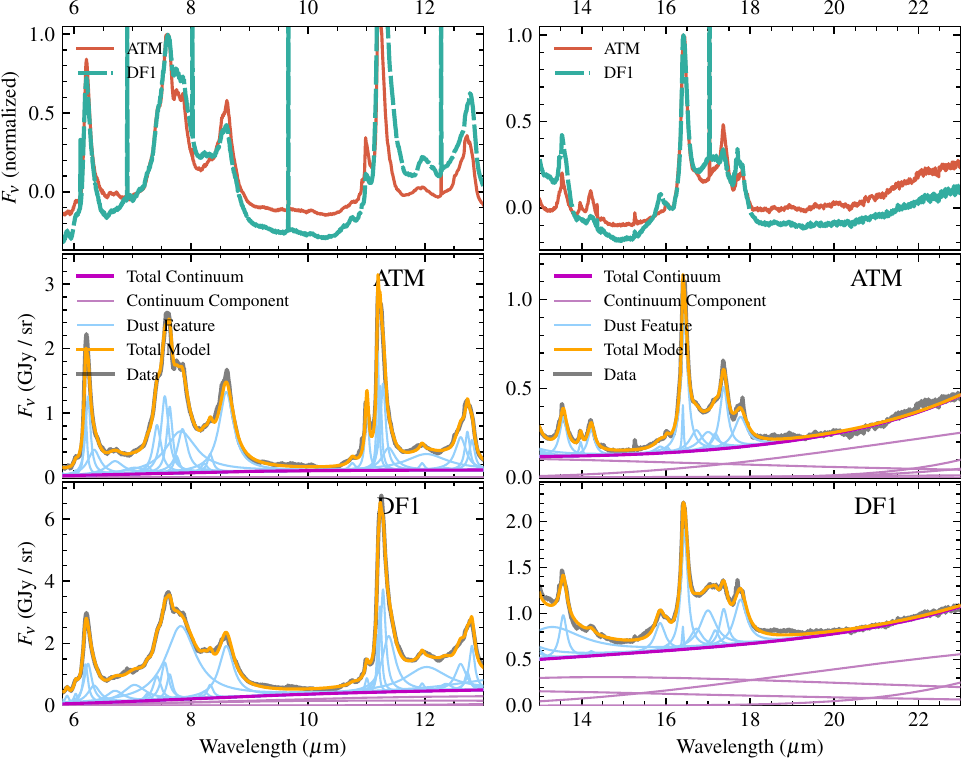}
    \caption{Analogous to Fig.~\ref{fig:specandfit}, for 6-13~\mum and 13-23~\mum. 
    The normalization in the top panels is based on the 7.7 and 16.4~\mum features respectively.
    The updated tuning for the 16-18~\mum complex better captures the shape changes, resulting in a more robust extraction of the power of the 16.4 and 17.4~\mum features.}
    \label{fig:specandfit2}
\end{figure*}

\subsection{Integrated feature power definitions}
\label{sec:totalpower}

\begin{table*}[bht]
    \centering
    \caption{Definitions used for total band power.}
    \label{tab:legend}
    \begin{tabular}{ccc}
        \hline\hline
        Band & Sub-components & Wavelengths\textsuperscript{a}  \\
        (Name) & (Names) & (\mum)\\
        \hline
        3.3 & A, B & 3.280, 3.295\\
        3.4 & A, B & 3.395, 3.403\\
        5.2 & A, B & 5.238, 5.289\\
        6.2 & A, B, C & 6.196, 6.239, 6.341\\
        7.7 narrow & 7.4, A, B, C, D & 7.418, 7.552, 7.635, 7.753, 7.854\\
        7.7 broad & broad & 7.823\\
        7.7 & narrow + broad & see above\\
        8.6 & -- & 8.604\\
        11.0 & -- & 11.008\\
        11.3 & A, B, C, D & 11.196, 11.233, 11.284, 11.381\\
        12.7 & A, B & 12.727, 12.804\\
        16.4 & A, B & 16.402, 16.439\\
        17.4 & -- & 17.374\\
        \hline
    \end{tabular}
    \tablefoot{\textsuperscript{a} Displayed wavelengths are taken directly from the updated PDR pack.}
\end{table*}

Most bands are decomposed into multiple sub-components by the PDR pack.
The PAHFIT model yields the power of each fitted Drude profile, integrated over the width of the profile in frequency space (surface brightness in W m$^{-2}$ sr$^{-1}$).
In the figures and sections that follow, we will often refer to the ``total integrated power'' for the major emission bands.
With this, we mean a sum of the aforementioned integrated Drude power, over all sub-components considered part of a band, as defined by \citet{VanDePutte2025}.
Considering the updates to the PDR pack presented in Sect.~\ref{sec:decompositionresults}, we provide an updated version of these sum definitions in Table~\ref{tab:legend}.

\subsection{Application of PAHFIT to the spectral data cube}
\label{sec:pahfitcube}

We fit each spaxel individually with PAHFIT, and the scripts that  perform the fits with parallel processing, and keep track of the input and output data, are provided as the PAHFITcube utility\footnote{\url{https://github.com/drvdputt/PAHFITcube.git}}.
Other notable features of this freely available tool include resuming a series of fits after an interruption, summarizing the results as spatial distribution maps, and plotting overviews of the maps similar to Fig.~\ref{fig:allmaps}.

\subsection{Limitations}
\label{sec:caveats}

This paper is an initial overview of the carbonaceous emission bands in our NGC\,7023 data.
Here we address a number of issues which may need to be addressed in followup work.
Since this work applies the PDR pack with PAHFIT, all the same caveats listed by \citet{VanDePutte2025} apply here as well, and here we highlight a few subjects which may require a more careful or dedicated analysis.

First, we note a few wavelength ranges where one should be careful when interpreting the power of the features as determined by these PAHFIT results.
For the weaker features in the 3.2-3.6~\mum range the fit is affected by the lack of description of the nebular continuum, as was noted in the `Limitations' section of  \citet{VanDePutte2025}.
For the 3.3~\mum band, it is notable that it does not seem to widen toward the blue side as was observed for the Orion Bar spectra, but this effect is subtle and a good continuum model is needed to quantify it.
For similar reasons, there is a degeneracy between the assumed FWHM of the broader features in the 3.8-4.7~\mum range (Table~\ref{tab:tune16}) and the assumed continuum level.
Typically, a continuum estimation that is too low will be compensated by a tuning with wider features.
The 6.0~\mum feature at the base of the blue wing of the 6.2~\mum band will be underestimated if there is a mismatch between the model and this blue wing.
For this work, the FWHM of the 8.6~\mum band was adjusted to 0.3~\mum, though this band also appears to have an additional narrow component, as can be seen in the ATM spectrum (Fig.~\ref{fig:specandfit2}), while the PDR pack approaches this with just one broad feature.
The 11.0 \mum feature is positioned on the steep blue wing of the 11.3 \mum band, and the 11.0 \mum feature itself also likely has a steep blue wing and a wider red wing.
For the Orion Bar, the 11.0~\mum profile and amplitude are sufficiently approximated by a single Drude profile \citep{VanDePutte2025}, but for NGC\,7023, this feature is much stronger relative to 11.3~\mum, and it seems multiple components are needed to reproduce the shape more closely \citep{Shannon2016}.
The smaller features redward of 13.5 (e.g.,~14.2~\mum) tend to be underestimated due to the influence of the wide component added to fit the red side of the 11-14 \mum complex (centered around 13.2 \mum), which is prominent in the DF1 spectrum.
Finally, the feature at 17.7 \mum appears rather complex, with a flat peak and some substructure, and is approximated by a single Drude profile.

To interpret the region beyond the DF1, the presence of ice absorption needs to be taken into account.
For the 3.3~\mum band, the H$_2$O absorption spanning 3.2-3.3~\mum \citep{Misselt2025} will have an effect, and additional ice absorption bands are found at 4.2-4.3~\mum (CO$_2$ stretching), at 14.8-15.4~\mum (CO$_2$ bending), and at 4.64-4.69~\mum (CO).
For this reason, this region \citep[MO1, DF2 and MO2 in][]{Misselt2025} is not the main focus of this work.
In addition, the presence of this ice indicates that a dust attenuation correction will be necessary to properly compare the inter-band ratios across the mosaic.

Finally, we note that this work does not make full use of the available resolution, since the data cubes were built on a common resolution matching MRS channel 4 ($0\farcs35$, see Sect.~\ref{sec:data}).
The phenomena we demonstrate are still clear with the larger spatial scales probed by our choice of spatial grid.
For future work that focuses on a specific wavelength range, data cubes that exploit the finer spatial resolution of the shorter channels could be built, and this will require point spread function (PSF) matching or interpolating appropriately.
Alternatively, the tools for fitting entire spectral cubes may need some form of regularization to address the difference in spatial resolution across the channels.

\section{Results}
\label{sec:results}

We focus our discussion on the brightest emission complexes with the strongest profile variations, and show
three views of the results throughout this section.
We start with a full-spectrum overview of the emission profile differences and how these are captured by the PAHFIT decomposition in Sects.~\ref{sec:mainbands} and \ref{sec:decompositionresults} (Figs.~\ref{fig:specandfit} and \ref{fig:specandfit2}).
We then discuss the spatial distributions of the main complexes (Fig.~\ref{fig:mapstotals}) and their sub-components (Figs.~\ref{fig:maps77}-\ref{fig:maps127}) in Sect.~\ref{sec:spatialdistribution}.

\subsection{Main bands}
\label{sec:mainbands}

Before discussing the spatial variations of the emission profiles and the corresponding spatial distributions, we provide a brief overview of common interpretations for each complex.
The main features in the NIRSpec range are the 3.3~\mum aromatic C-H stretching and 3.4~\mum aliphatic C-H stretching features (Fig.~\ref{fig:specandfit}, left panels).
As already shown by \citet{Misselt2025}, the 3.4/3.3~\mum ratio that traces the aliphatic material abundance is higher for the DF, analogous to what a similar decomposition reveals for the Orion Bar \citep{Peeters2024}.
Two weaker features between 3.8 and 4.8~\mum (Fig.~\ref{fig:specandfit}, center panels) were attributed to analogous stretching modes of deuterium-substituted material: the deuterated 4.4~\mum aromatic and 4.7~\mum aliphatic features \citep{Misselt2025}.
The identification of the weaker features at 5.2, 5.7, 5.9, and 6.0~\mum (Fig.~\ref{fig:specandfit}, right panels) is not as well established.
Previously, \citet{Chown2024} and \citet{VanDePutte2025} identified and quantified the shape changes of the 5.7~\mum feature between the Atomic PDR and DF regions of the Orion Bar.
Through overtones or combination modes, a connection was proposed with the aromatic C-H out-of-plane bending modes in the 10-15~\mum range \citep{Allamandola1989, Boersma2009, Mackie2015, Chown2024}.

The wavelength range shown in Fig.~\ref{fig:specandfit2} contains a number of major complexes.
The bands near 6.2~\mum and 8.6~\mum are typically enhanced for PAH cations, and hence considered as tracers of cationic aromatic species \citep{Allamandola1999, Peeters2002}.
The 6.2~\mum band is associated with aromatic C-C stretching, and the 8.6~\mum band with aromatic C-H in-plane bending.
The 7.7~\mum complex, exhibits narrow bands attributed to mixed-character aromatic C-C stretching or C-H in-plane bending modes, also associated with cations.
These bands appear to be perched on top of a broader component near 7.9~\mum, which has previously been associated with the presence of PAH clusters or very small grains (VSGs; \citealt{Berne2007, Pilleri2012, Peeters2017, Stock2017, Khan2025}).
The decomposition with the PDR pack attempts to separate these narrow and broad contributions, and the broad-to-total ratio determined based on the PAHFIT results was shown to increase with the distance from the star in the Orion Bar \citep{VanDePutte2025}.

The 10-15~\mum range, contains various aromatic C-H out-of-plane bending modes, with the strongest peaks near 11.3 and 12.7~\mum.
The total 11.3~\mum flux is often used in ratios such as 6.2/11.3~\mum to trace neutral PAHs, while there are components associated with cationic species as well, such as the 11.0~\mum feature \citep[e.g.,][]{Hudgins1999} and a blue-side component of the 11.3~\mum band identified as 11.207~\mum by \citet{Khan2025}.
We note that Fig.~\ref{fig:specandfit2} shows that the 11.0~\mum feature is significantly stronger in the ATM spectrum.
These modes can also be interpreted in terms of hydrogen adjacency at the edges of the aromatic material \citep[solo, duo, trio, quartet; e.g.,][]{Hony2001, Bauschlicher2008, Khan2025}.
For example, the 11.3~\mum band is dominated by solo hydrogens, and the 12.7~\mum band by duo and trio hydrogens.

The group of bands in the 16-18~\mum range consists of aromatic C-C-C bending modes, also known as skeletal modes \citep{Allamandola1989, VanKerckhoven2000}.
These likely contain information about the structure on a larger scale, since they involve the bending of larger groups of atoms \citep{Peeters2004, Boersma2010}, and the ratios of the components have connections to the charge state as well \citep{Peeters2012, Shannon2015}.
The spectra in the top right panel of Fig.~\ref{fig:specandfit2} are normalized to the 16.4~\mum band, which emphasizes the difference in the relative 17.4~\mum feature strength.

\subsection{Decomposition and PDR pack updates}
\label{sec:decompositionresults}

To emphasize the profile variations, extracted spectra for the ATM and DF1 regions are shown in the top panels of Figs.~\ref{fig:specandfit} and \ref{fig:specandfit2}, normalized by applying a vertical shift and a normalization factor using reference points at the peak and the base of a particular feature (as stated in each caption).
Below we describe a number of updates that were made to the PDR pack, based on the details and differences between ATM and DF1 that are revealed by this comparison.
A detailed description of the decomposition as specified by the PDR pack is provided in the appendix of \citet{VanDePutte2025}, and will not be repeated here.

\begin{figure}[tbh]
    \centering
    \includegraphics[]{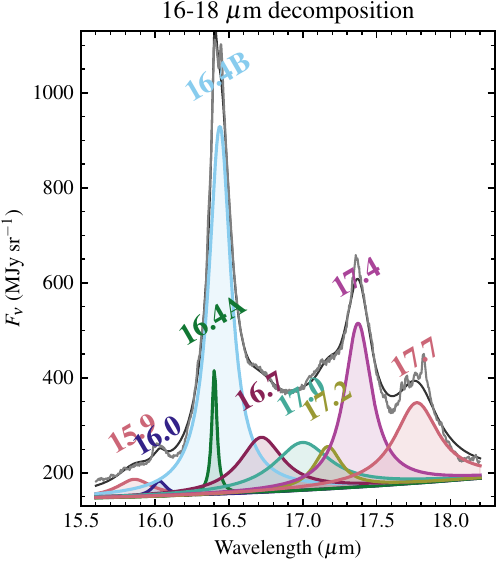}
    \caption{Detailed view of the updated decomposition in the 16-18~\mum range.
    The light gray line shows the ATM spectrum, with bumps at 16.7 and 17.2~\mum that raise the need for three components to model the central ``plateau''.
    The individual Drude profiles (colorful curves and labels) are plotted after adding the continuum components, and the total model is the solid black line.}
    \label{fig:legend170}
\end{figure}

The 16.4 \mum profile is similarly shaped for the ATM and DF1 spectra of NGC\,7023, while the rest of the 16-18~\mum complex exhibits significant differences.
Compared to the Orion Bar spectra of \citet{VanDePutte2025}, these differences are more strongly pronounced in NGC\,7023 owing to its fainter continuum.
For this reason, the decomposition by the original PDR pack could not sufficiently capture the flux of the 17.4~\mum band and the details of the emission profile variations.
We therefore updated the PDR pack for this wavelength range using a procedure analogous to \citet{VanDePutte2025}.
A set of central wavelengths and FWHM values were determined for both the ATM and DF1 spectrum by applying a least-square fit to the 15.7-18.0~\mum range, and a continuum subtraction based on a preliminary PAHFIT model.
The results for this update are summarized in Table~\ref{tab:tune16}.
The main difference is that the total number of Drude profiles has increased by 2, to better describe the shape of the plateau between the 16.4 and 17.4~\mum bands.
The new features were motivated based on two bumps seen in the ATM spectrum (Fig.~\ref{fig:legend170}): one on the red side of the 16.4~\mum feature, and one near the blue side of the 17.4~\mum feature.
In the DF1 spectrum, the plateau peaks near 17.0~\mum instead, indicating a third feature.
The version of the PDR pack based on the Orion Bar data could not sufficiently capture these details, as it defined only two wide components centered at 16.8 and 17.1 \mum to model this plateau.

These data also provide an opportunity to determine a set of Drude profiles near 3.8-5.0~\mum for the PDR pack.
This wavelength range is problematic in the Orion Bar data, because of the wavelength gap in the G395H observation and the larger number of emission lines.
Our tuning for the features near 3.8, 4.0, 4.35, and 4.75~\mum is given in Table~\ref{tab:tune16}.
We note that the features in Table~\ref{tab:tune16} are likely too wide, as their tuning partially compensates for the lack of a continuum description
\citep[see `Limitations' section of][]{VanDePutte2025}.
A realistic continuum description is challenging as there are possible contributions from nebular emission or PAH-related quasi-continua \citep{Allamandola1989, Boersma2023}.

The updated PDR pack was applied to the full 3.2-26~\mum range, and two resulting fits are shown in the lower two panels of Figs.~\ref{fig:specandfit} and \ref{fig:specandfit2}.
As stated in Sect.~\ref{sec:decompositionmethod}, we do not use the PAHFIT dust attenuation model for the final results.
As a test, we performed the decomposition both with and without the classic PAHFIT dust attenuation model \citep{Smith2007}, and the attenuation resulting from the fit was zero for both the ATM and DF1 spectra.
We note that \citet{Fleming2010} applied IDL PAHFIT to Spitzer spectra of NGC\,7023, and found a negligible optical depth at the peak of the silicate extinction feature at 9.7 \mum ($\tau_{9.7} < 0.2$).
Past the DF, some level of extinction is expected, considering that ice absorption is seen in the NIRSpec data \citep{Misselt2025}.
Therefore, as stated in Sect.~\ref{sec:caveats}, this region is not the main focus in this work.

\begin{table}[bth]
    \caption{PDR pack update based on NGC\,7023.}
    \label{tab:tune16}
            \begin{center}
    \begin{tabular}{lcc}
    \hline
    \hline
    Name & Wavelength & FWHM \\
    & (\mum) & (\mum) \\
    \hline
    \multicolumn{3}{c}{Update\ for 3.8-4.8 \mum}\\
    3.8 & 3.79 & 0.40 \\
    4.0 & 4.09 & 0.19 \\
    4.35 & 4.35 & 0.34 \\
    4.75\textsuperscript{a} & 4.743 & 0.1\\
    4.75 broad\textsuperscript{a} & 4.743 & 0.35\\
    \multicolumn{3}{c}{Update for 16-18 \mum}\\
    16.0A & 15.86 & 0.289 \\
    16.0B & 16.03 & 0.114 \\
    16.4A & 16.402 & 0.040 \\
    16.4B & 16.439 & 0.182 \\
    16.7\textsuperscript{b} & 16.720 & 0.352 \\
    17.0 broad\textsuperscript{b} & 16.997 & 0.464 \\
    17.2\textsuperscript{b} & 17.170 & 0.219 \\
    17.4 & 17.374 & 0.224 \\
    17.7 & 17.771 & 0.362 \\
    \hline
    \end{tabular}
    \end{center}
    \tablefoot{%
    \textsuperscript{a}The approximation in \citet{VanDePutte2025} consisted of a single, much broader 4.75~\mum feature (FWHM $\sim 0.9$~\mum).
    \textsuperscript{b}These three features replace ``16.765'' and ``17.112'' from \citet{VanDePutte2025}}
\end{table}

\subsection{Spatial distributions}
\label{sec:spatialdistribution}

\begin{figure}[tb]
    \centering
    \includegraphics[]{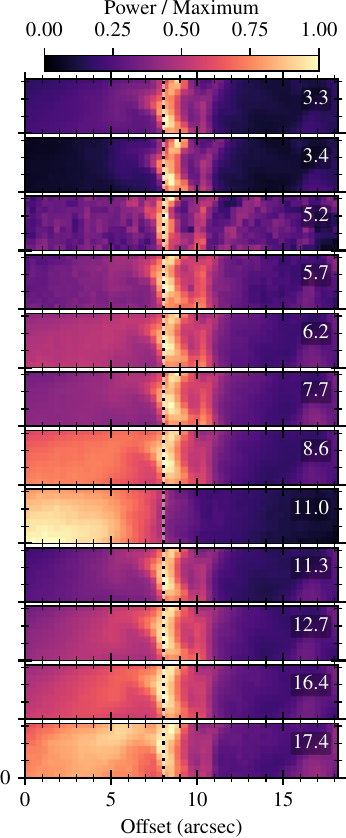}
    \caption{Power maps of the main bands based on sums of the sub-components contributing to each band as defined in Table~\ref{tab:legend}.
    The color scale represents the spectrally integrated power (W m$^{-2}$ sr$^{-1}$) normalized to the brightest spaxel.
    The dashed black vertical line corresponds to DF1 (more precisely: the maximum of 16.4~\mum when averaged vertically), and the ATM region is in the leftmost 3'' of the image.}
    \label{fig:mapstotals}
\end{figure}

A selection of spatial distributions resulting from our method (Sect.~\ref{sec:pahfitcube}) is displayed in Figs.~\ref{fig:mapstotals} through~\ref{fig:maps127}.
A complete overview of maps for all individual PDR pack components is provided in Appendix~\ref{app:allmaps}.
We limit the scope of the discussion to the bands at 3.3, 3.4, 5.2, 5.7, 6.2, 7.7, 8.6, 11.0, 11.3, 12.7, 16.4 and 17.4~\mum.
While the PDR pack also fits numerous weaker features, those fits are more sensitive to being biased by a mismatch in the continuum model, and such features require an individualized extraction approach.

\begin{figure}
    \centering
    \includegraphics{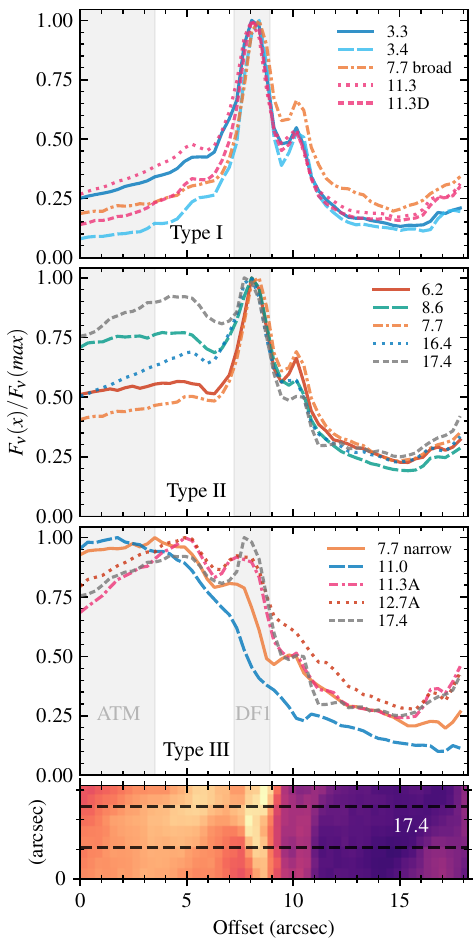}
    \caption{Spatial profiles along cuts perpendicular to the DF for selected features, normalized to the peak of each cut.
    The cuts are organized by spatial distribution types I, II, and III, as introduced in Sect.~\ref{sec:spatialdistribution}.
    The 17.4~\mum cut is shown twice to directly compare it to the type II and type III cuts.
    The dashed lines in the bottom panel indicate the region over which the maps were median-stacked along the vertical axis to obtain the cuts, with the 17.4~\mum map from Fig.~\ref{fig:mapstotals} in the background.
    The shaded areas indicate the widths of the ATM and DF1 regions.}
    \label{fig:cuts}
\end{figure}

One of the main points of this discussion, is how individual sub-components have spatial distributions that differ from the integrated power in each band (Sect.~\ref{sec:totalpower}, Table~\ref{tab:legend}).
We start with an overview of the total power maps in this section.
In what follows for the 5.7, 7.7, 11.3, and 12.7~\mum bands, Figs.~\ref{fig:maps57}, \ref{fig:maps77}, \ref{fig:maps112}, and \ref{fig:maps127} zoom in on the sub-components and show maps of the individual components alongside the corresponding total (see Sect.~\ref{sec:subcomponents}).
In the following sections, we describe the similarities and differences observed between these spatial distributions, further clarified by the one-dimensional cuts shown in Fig.~\ref{fig:cuts}.
For each cut, the data point are medians over the vertical range shown in the bottom panel.
We set the lower limit of the vertical axis to zero, to emphasize that all emission maps are non-zero on both sides of DF1, and normalize each one-dimensional spatial profile to its peak brightness.

\begin{figure}[tbh]
    \raggedleft
    \includegraphics[scale=1]{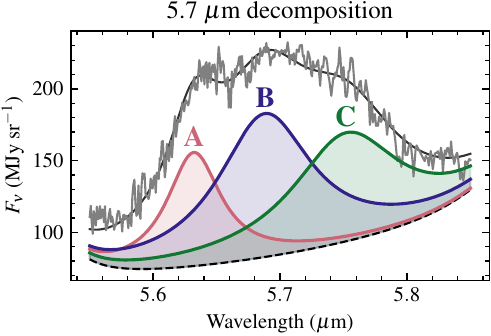}
    \includegraphics[width=0.82\linewidth]{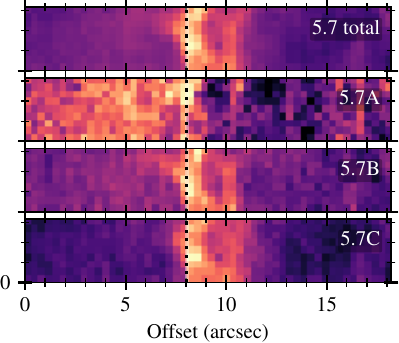}
    \caption{Maps of the individual components of the 5.7~\mum band decomposition.
    The lower panel is analogous to Fig.~\ref{fig:mapstotals}.
    The top panel is a legend illustrating the sub-components labeled A, B, and C.
    The light gray line shows the ATM spectrum data, the solid black line is the best fitting total model, and the dashed line shows the sum of all other components (continuum and wings of nearby Drude profiles), to which the three 5.7~\mum components are added.}
    \label{fig:maps57}
\end{figure}

\begin{figure}
    \raggedleft
    \includegraphics[scale=1]{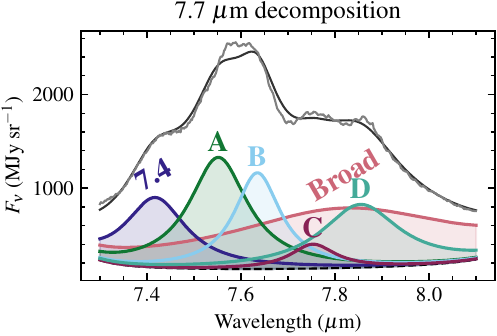}
    \includegraphics[width=0.82\linewidth]{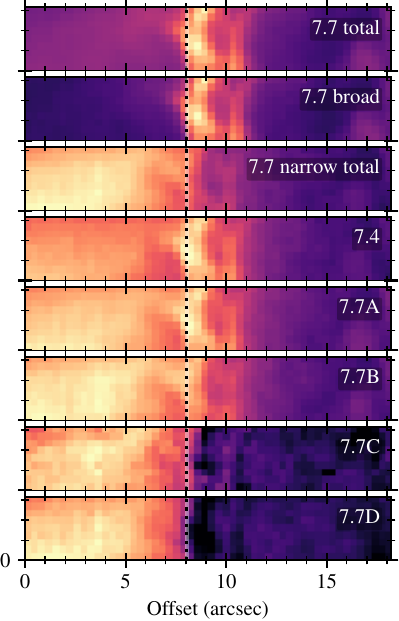}
    \caption{Maps of the 7.7~\mum decomposition and legend analogous to Fig.~\ref{fig:maps57}.
    The sum of the narrow components (as in Table~\ref{tab:legend}) is shown in addition to the individual maps.}
    \label{fig:maps77}
\end{figure}

\begin{figure}[tbh]
    \raggedleft
    \includegraphics[scale=1]{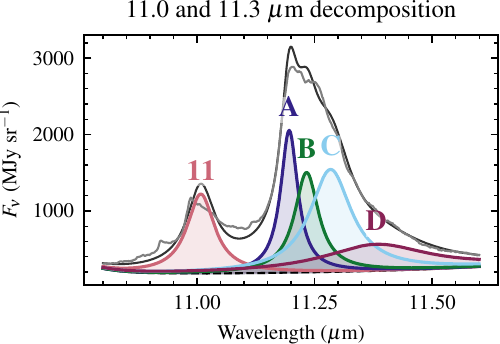}
    \includegraphics[width=0.82\linewidth]{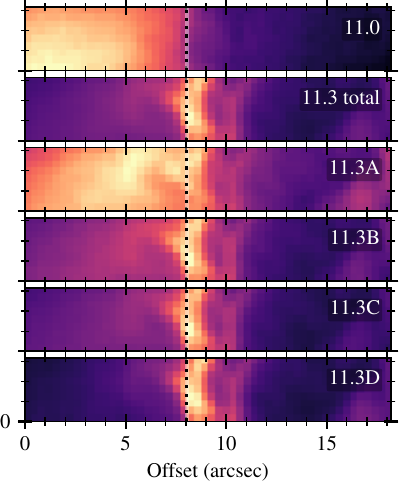}
    \caption{Analogous to Fig.~\ref{fig:maps57} for the 11.0 and 11.3~\mum bands.}
    \label{fig:maps112}
\end{figure}

\begin{figure}
    \raggedleft
    \includegraphics[scale=1]{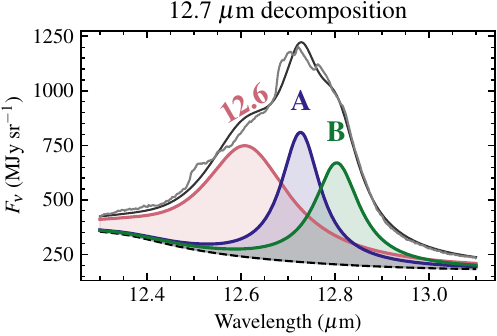}
    \includegraphics[width=0.82\linewidth]{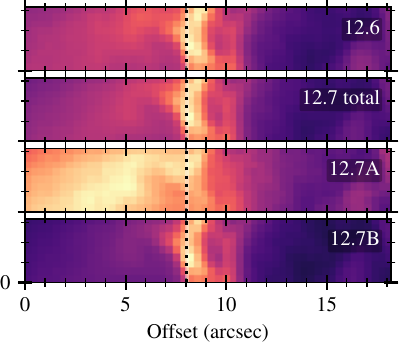}
    \caption{Analogous to Fig.~\ref{fig:maps57} for the 12.7~\mum band.
    Note the similarity of 12.7A to 11.3A in Fig.~\ref{fig:maps112}.}
    \label{fig:maps127}
\end{figure}

\subsubsection{High and low emission in the ATM region}
\label{sec:types}

A common element among all maps in Fig.\ref{fig:mapstotals} is a maximum at DF1, with the 11.0~\mum band being the only exception.
In the figures that follow as well as the appendix (Fig.~\ref{fig:allmaps}), several other maps lacking the DF1 peak are seen: the smooth decrease of 11.0~\mum with no trace of the DF1 peak also appears for 10.8~\mum (Fig.~\ref{fig:allmaps}).
Other maps lacking the DF1 peak do not show the same smooth behavior; they are noisier (e.g., 5.44), more biased by the continuum fit (e.g., 14.0), or suffer from degeneracy with another feature (e.g., 7.7C, 7.7D), and show a region with zero power near DF1 as a result.

The most conspicuous difference between the maps is found in the star-facing half of the field of view (which contains ATM): a wide variety of emission strengths relative to the DF1 peak.
Hence, we categorized the features into three types based on their relative brightness in the ATM region compared to their brightness at DF1.
The ratio $P_\text{ATM}/ P_\text{DF1}$ can be low ($< 0.5$, type I), intermediate ($> 0.5$, $< 0.8$ type II), or high ($> 0.8$, type III).
While there are more details that characterize the spatial profiles, this grouping into types based on the ATM/DF1 intensity ratio will aid the discussion.
Considering the total band emission shown in Fig.~\ref{fig:mapstotals}, and the cuts in Fig.~\ref{fig:cuts},
the bands at 3.3, 3.4, 5.2, 5.7, 7.7, and 11.3~\mum are of type I, those at
6.2, 7.7, 8.6, 12.7, and 16.4~\mum are of type II,
while the 11.0 and 17.4~\mum features are of type III.
Spatial cuts for several type I features are grouped in the top panel of Fig.~\ref{fig:cuts}.
The 3.3 and 11.3~\mum (total) cuts are nearly identical, and have matching peak positions despite being observed with different instruments.
The center panel shows several type II spatial profiles.
For the spatial cuts of 6.2 and 8.6~\mum, we note the parallel nature of the curves on the ATM side, despite the intensity difference in that region.

We emphasize the difference between the 16.4 and 17.4~\mum bands.
The 17.4~\mum cut is shown in both the type II and type III panels of Fig.~\ref{fig:cuts} to facilitate a comparison to the 16.4~\mum cut.
The 16.4 and 17.4~\mum cuts have matching peak positions and behave similarly beyond DF1.
The spatial profiles are nearly parallel in the ATM region as well, though the 17.4~\mum feature exhibits a secondary maximum nearly as strong as the DF1 peak, while the strength of 16.4~\mum is suppressed relative to DF1.
The nature of these two features is discussed further in Sect.~\ref{sec:discuss164}.

\subsubsection{Individual sub-components}
\label{sec:subcomponents}

The top panels of Figs.~\ref{fig:specandfit} and \ref{fig:specandfit2} show the band profile differences between ATM and DF1, and these variations are similar to those seen in the Orion Bar \citep{Chown2024}.
The fits with the PDR pack capture some of these profile differences, if the relative contributions of the sub-components to each band are considered.
The maps of the individual sub-components of the 5.7, 7.7, 11.3, and 12.7~\mum complexes (Figs.~\ref{fig:maps57}-\ref{fig:maps127}) reveal additional spatial distributions of type III (bright in ATM).
These figures include a panel with a detailed view of the decomposition, showing the ATM spectrum and its PAHFIT model to clarify the interpretation of the sub-component maps.

For the 7.7~\mum maps (Fig.~\ref{fig:maps77}), we have added maps of two partial power sums: ``7.7 narrow'' and ``7.9 broad'' (Table~\ref{tab:legend}, Sect.~\ref{sec:totalpower}).
A spatial cut of the ``7.7 narrow'' combination is shown in the lower (type III) panel of Fig.~\ref{fig:cuts}, and it is remarkably similar to the 11.0~\mum feature.
As noted in the previous section, the 11.0~\mum cut has less substructure; there is no peak at DF1.
The top panel of Fig.~\ref{fig:maps112} shows that the 11.0~\mum feature is not very well approximated by a single Drude profile, making it likely that the 11.0~\mum profile has a pronounced red tail similar to 11.3~\mum.
We verified the 11.0~\mum spatial distribution result by applying an ad-hoc 11.0~\mum measurement that consists of fitting a local linear continuum and numerically integrating the data.
This approach results in a nearly identical spatial distribution.

\begin{figure}
    \centering
    \includegraphics{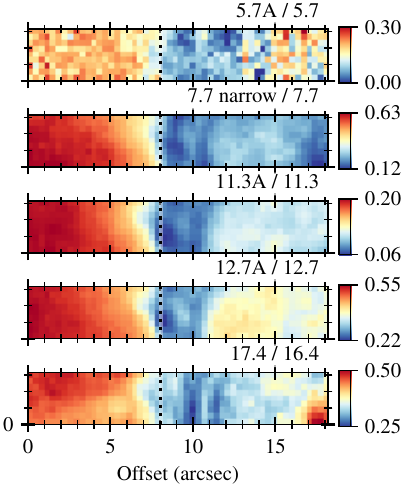}
    \caption{
    Maps of ratios describing variations in the band profiles.
    The color bars represent a linear scale between the minimum and maximum.
    The ratios represent the relative contributions of sub-components to the total as defined in Table~\ref{tab:legend}.
    The ratios of the bluest components of each complex are shown alongside the 17.4 / 16.4 ratio to illustrate their similarity, and how the steepest gradient is in the same location.}
    \label{fig:ratiomaps}
\end{figure}

The 11.3 and 12.7~\mum bands both exhibit profile changes, and the PDR pack explains this as their bluest sub-components being of type III.
The 11.3~\mum band is described by four sub-components (A, B, C, and D) all of which are shown in Fig.~\ref{fig:maps112}.
The peak of the 12.7~\mum band is decomposed into two features, although the blue wing contains a significant contribution by the 12.6~\mum component (Fig.~\ref{fig:maps127}).
We show cuts for the bluest sub-components called 11.3A and 12.7A (see also Table~\ref{tab:legend}) in the type III panel of Fig.~\ref{fig:cuts}, showing the similarity with the 17.4~\mum band.
Analogous behavior seems to be present for the blue component of the 5.7~\mum band, though the spectra for this band features and the resulting maps of the components are noisier (Fig.~\ref{fig:maps57}).
The cuts are also compared to 11.0~\mum, and we note how 11.3A and 12.7A still have more substructure than 11.0~\mum.
Therefore, 11.0~\mum could be considered a special or extreme case of the type III spatial distribution.

Conversely the reddest sub-components tend to exhibit type I spatial distributions, being suppressed in the ATM region and having a strong peak at DF1.
The clearest examples are 5.7C (Fig.~\ref{fig:maps57}), 7.7 broad (Fig.~\ref{fig:maps77}), 11.3D (Fig.~\ref{fig:maps112}) and 12.7B (Fig.~\ref{fig:maps127}).
The 7.7 broad spatial cut was added to the type I (top) panel of Fig.~\ref{fig:cuts}, showing its similarity to the 3.4~\mum aliphatic C-H feature, with nearly identical peak positions and widths.
To further emphasize the similarity of the spatial behavior of the profile variations at 5.7, 7.7, 11.3, and 12.7~\mum, we show sub-component-to-total ratio maps (e.g., 5.7A / 5.7) in Fig.~\ref{fig:ratiomaps}.
We note how these values show a positive gradient (i.e., the profiles become more blue-side dominated) in the direction toward the star, that persists up to the left edge of the field of view.
We propose by extrapolation that the profiles keep evolving beyond this boundary, for regions progressively closer to the star (see also Sect.~\ref{sec:outlook}).

Summarizing, the sub-component maps reveal that the bluest sub-components at 5.7, 7.7, 11.3, and 12.7~\mum have type III spatial distributions, even though the total band emission is of type I or type II.
We provide an overview of the spatial distribution types of the features in Table~\ref{tab:types}, together with the $P_\text{ATM}/ P_\text{DF1}$ ratio, in which the feature strengths $P_\text{ATM}$ and $P_\text{DF1}$ are measured based on the fits shown in Figs.~\ref{fig:specandfit} and \ref{fig:specandfit2}, and the strengths of summed features are computed as defined in Table~\ref{tab:legend}.

\begin{table}[bth]
    \caption{Spatial distribution types and brightness ratios for the main features and sub-components.}
    \label{tab:types}
    \centering
    \begin{tabular}{lll|c}
        \hline\hline
        \multicolumn{3}{c|}{Name and type} & Ratio\\
        I & II & III & $P_\text{ATM}/ P_\text{DF1}$\\
        \hline
        3.3 &&& 0.36 \\
        3.4 &&& 0.14\\
        5.2 &&& 0.40\\
        \textbf{5.7} &&& 0.40\\
        & 5.7A && 0.66\\
        5.7B &&& 0.44 \\
        5.7C &&& 0.30 \\
        & 6.2 && 0.59\\
        6.2A &&& 0.45\textsuperscript{a}\\
        & 6.2B && 0.60\textsuperscript{a}\\
        & \textbf{7.7} && 0.50 \\
        7.7 broad &&& 0.27\\
        && 7.7 narrow & 1.57 \\
        && 7.4 & 0.96 \\
        && 7.7A & 0.99 \\
        && 7.7B & 1.15 \\
        && 7.7C & -\textsuperscript{b} \\
        && 7.7D & -\textsuperscript{b} \\
        & 8.6 && 0.75 \\
        && 11.0 & 2.26\textsuperscript{c}\\
        \textbf{11.3} &&& 0.37\\
        && 11.3A & 1.00\\
        11.3B &&& 0.47\\
        11.3C &&& 0.41\\
        11.3D &&& 0.19\\
        & 12.6 && 0.60 \\
        & \textbf{12.7} && 0.51 \\
        && 12.7A & 1.01\\
        12.7B &&& 0.33\\
        & 16.4 && 0.61\\
        && 17.4 & 0.85\\
        \hline
    \end{tabular}
    \tablefoot{
        \textsuperscript{a}The fitted 6.2~\mum components have variable FWHM. See also Sect.~\ref{sec:compareorion} and \citet{VanDePutte2025}.
        \textsuperscript{b} Zero in DF1 due to fitting degeneracy with the 7.7 broad component.
        \textsuperscript{c} 11.0~\mum can be considered a special case as it decreases smoothly without a local maximum at DF1.}
\end{table}

\section{Discussion}
\label{sec:discussion}

There are numerous previous studies that mapped the carbonaceous emission bands in NGC\,7023 using Spitzer data \citep{Werner2004}, cited throughout the discussion that follows.
With the typical $1\farcs8$ resolution of Spitzer-based spectral maps, details such as the structure and the precise location of the DF could not be resolved.
The spatial resolution of our JWST-based maps reveals that most of the changes in the carbonaceous band characteristics occur near DF1 or at around $3\arcsec$ ahead of this DF (see ratio maps of Fig.~\ref{fig:ratiomaps}).
In the following sections, we start by discussing the differences between the spatial distributions of the 16.4 and 17.4~\mum features.
This is followed by a discussion of possible relations with tracers based on the other bands and interpretations from Spitzer-based work.

\subsection{Spatial distribution of 16.4 and 17.4 \mum}
\label{sec:discuss164}

Originally, \citet{Sellgren2010} reported that the 16.4 and 17.4~\mum spatial distributions were similar in NGC\,7023.
In later works, it is typically indicated that two spatial distributions exist for this emission complex, one of which extends further toward the star \citep{Rosenberg2011, Boersma2013, Boersma2014, Boersma2015, Shannon2015, Sidhu2023}.
For example, \citet{Shannon2015} discussed the morphology of  the 16.4 and 17.4~\mum features, and noted how maps of the 17.4 and 11.0 \mum emission are both brighter close to the star, and that maps of the 12.7 and 16.4~\mum emission are similar as well.
Our maps and cuts presented in Sect.~\ref{sec:spatialdistribution} clarify that the 16.4 and 17.4~\mum bands have highly similar spatial distributions on the far side of the DF, while in the direction of the ATM region, the 17.4~\mum emission is enhanced (Figs.~\ref{fig:mapstotals} and~\ref{fig:cuts}).
The high resolution of JWST reveals that a change in the 17.4/16.4 ratio takes place near the boundary between DF1 and the more irradiated atomic PDR region (Fig.~\ref{fig:ratiomaps}).
It is therefore likely that UV processing plays a role, where the resulting changes to the emitting population (molecular size, charge, or structure) lead to different contributions by the relevant vibrational modes.

As we introduced in Sect.~\ref{sec:mainbands}, the vibrational modes linked to these bands are  skeletal C-C-C modes, as revealed by computed spectra of large PAHs \citep{Allamandola1989, VanKerckhoven2000}.
More specifically, the computational results of \citet{Ricca2010} show contributions at 16.4~\mum by in-plane elongation/compression modes, and contributions to 17.4~\mum by modes involving C-H out-of-plane bending.
Based on spectra in the NASA Ames PAH database (PAHdb), the presence of pendent rings was suggested as an explanation for the 16.4~\mum band \citep{Boersma2010}, while larger but compact PAHs \citep{Ricca2012} in the 50-130 carbon atoms range consistently show an emission band near 17.4~\mum.

\citet{Peeters2012} organized the components contributing to the 16-18~\mum emission into three groups, based on the correlation between the bands as observed in NGC\,2023.
The 16-18~\mum plateau and 15.8~\mum band correspond best to the 11.3~\mum emission, the 16.4~\mum band closely matches the 12.7~\mum emission, and the 17.4~\mum emission appears to represent a more spatially diffuse component.
\citet{Peeters2012} associate the 17.4~\mum band to doubly ionized PAHs or a certain subset of cationic PAHs.
Previously, it was concluded that the (normalized) 16-18~\mum bands are mostly uncorrelated among themselves, and that (besides the 12.7-16.4~\mum connection) there is no obvious correlation with the MIR features at 7-9~\mum and 11-15~\mum \citep[e.g.,][]{Boersma2010}.
The JWST spectra, our decomposition, and the spatial distributions of the individual sub-components reveal otherwise (Sect.~\ref{sec:subcomponents}).
In particular, it appears that the 17.4~\mum spatial distribution is particularly similar to the blue-side and narrower components of the 7.7, 11.3, and 12.7~\mum bands, as we demonstrate via correlation diagrams in the next section.

\subsection{Correlations of 16.4 and 17.4~$\mu$m with feature ratios}
\label{sec:corr}

\begin{figure*}[tbhp]
\centering
    \includegraphics[height=9cm]{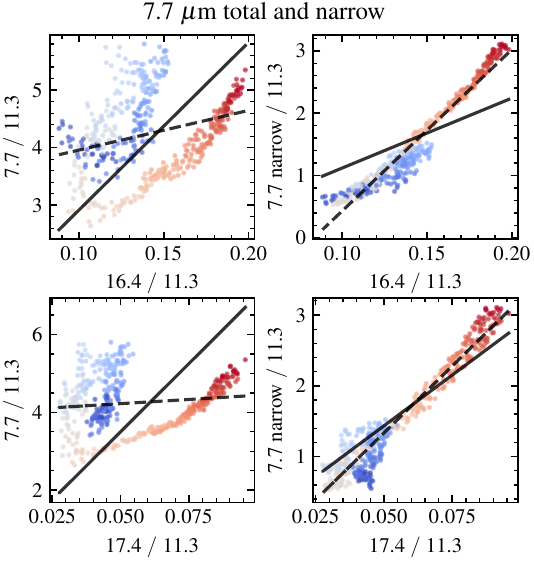}
    \includegraphics[height=9cm]{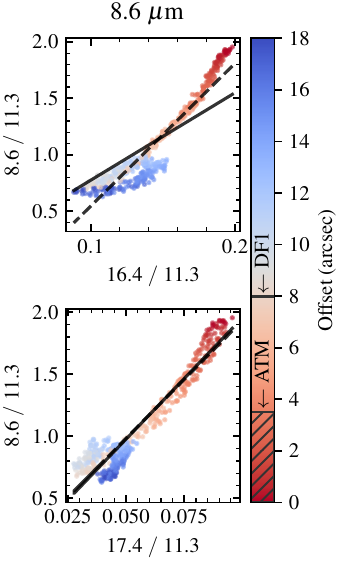}
    \includegraphics[height=9cm]{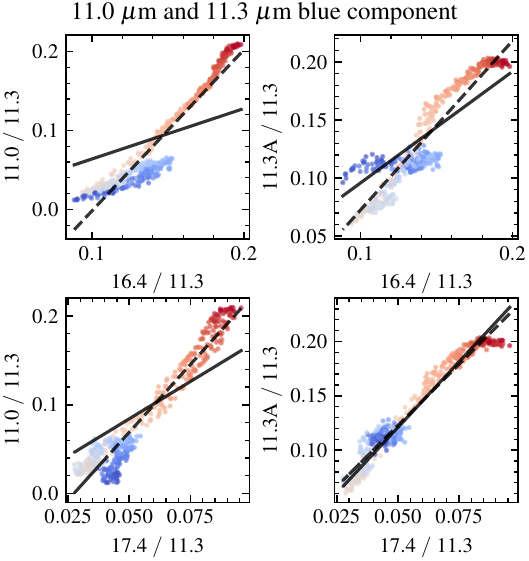}
    \includegraphics[height=9cm]{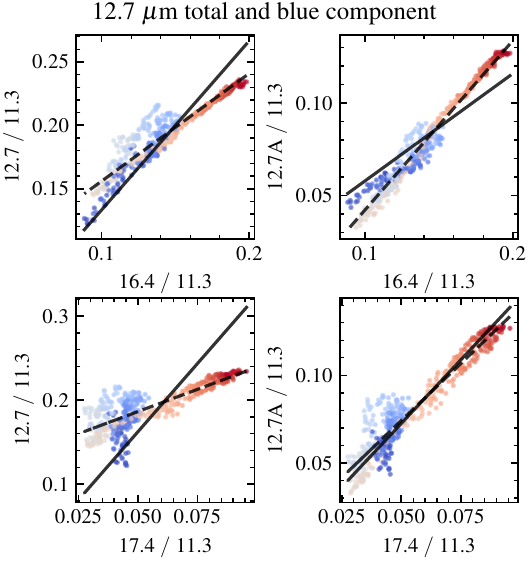}
    \caption{Correlation plots exploring the relation between the 16.4 and 17.4~\mum bands and a selection of other bands.
    The data points represent the spaxels of the maps, and their colors represent the offset from the edge of the mosaic, with the locations of ATM and DF1 indicated on the color bar.
    The black lines show linear fits through the origin ($y = ax$, solid) and with a variable intercept ($y = bx + c$, dashed).
    These lines are typically closer for 17.4~\mum, and more separated for 16.4~\mum, with the exception of the 12.7~\mum panels.
    This indicates that the 7.7 narrow, 8.6, 11.3A, and 12.7A features relate better with 17.4~\mum, while the total 12.7~\mum flux relates beter with 16.4~\mum (Sect.~\ref{sec:corr}).
    }
    \label{fig:corr}
\end{figure*}

The previous section mentioned that it was unclear what features best correlate with 17.4~\mum \citep{Boersma2010}, while the spatial position of the peak emission in NGC\,2023 indicates that the 17.4~\mum band is grouped with the 8.6 and 11.0~\mum bands, and the 16.4~\mum band with the 6.2, 7.7 and 12.7~\mum bands \citep{Peeters2017}.
Our results in Sect.~\ref{sec:spatialdistribution} revealed a visual similarity between the spatial distributions of the 17.4~\mum band and the narrower bluer sub-components of the 7.7, 11.3 and 12.7~\mum bands, as well as the 11.0~\mum band.

In this section we discuss correlation diagrams (Fig.~\ref{fig:corr}) comparing the 16.4 and 17.4~\mum bands to the aforementioned bands and their blue sub-components.
All quantities shown are normalized to the total 11.3~\mum band.
While not discussed earlier (see Sect.~\ref{sec:caveats}), we included the molecular region (``Mol'') in the diagrams for completeness (on the right side of DF1 in the maps).
As a visual aid, we added two linear fits to each panel of Fig.~\ref{fig:corr}: one which is a simple scaling relation through the origin ($y = ax$), and one which allows a non-zero intercept ($y = bx + c$).
When these linear fits are parallel, then extrapolating the relationship predicts a behavior where both quantities approach zero simultaneously, indicating that the emitting species are co-spatial and behave as a single population with no other contributions.
Conversely, a larger separation between the linear fits visualizes a non-zero intercept.
Even if the correlation is strong, this means that at least one of the two bands in the diagram has significant contributions from another population with a different spatial distribution.

For the total 8.6, 11.0 and 12.7~\mum bands, we find a result that is consistent with the grouping of \citet{Peeters2017}:
the 17.4~\mum feature relates best to the 8.6 and 11.0~\mum features, while the 16.4~\mum band appears to relate best to the total of 12.7~\mum.
We conclude this from the linear fits of Fig.~\ref{fig:corr}: in each quadrant, compare the lines for 16.4~\mum (top row) and 17.4~\mum (bottom row).
In most cases, the lines are closer together for the 17.4~\mum diagram, while the total 12.7~\mum band is the exception, establishing its relation with the 16.4~\mum band instead.
The total 7.7/11.3 ratio shows a strong bifurcation in the diagram, because of the varying contribution by the 7.7 broad component, which gets stronger on the shielded side of the DF.
This becomes even more apparent if the molecular region is included (blue points).
This indicates that both the 16.4 and 17.4~\mum features are likely not related to the broad 7.7~\mum component.

Inspecting the sub-components at 11.3 and 12.7~\mum reveals additional relationships with the 17.4~\mum band.
While the total 11.3~\mum emission cannot be shown as it is the normalization, we can still show the blue sub-component ratio 11.3A/11.3 here, which describes the band profile.
Large 11.3A values correspond to a sharper profile, similar to that of the ``atomic PDR'' template spectrum of the Orion Bar \citep{Chown2024}.
There is some difference between the two linear fits for 16.4~\mum, while the lines almost overlap for 17.4~\mum.
This shows that while both 16.4 and 17.4~\mum correlate strongly with 11.3A and 12.7A, the 17.4~\mum band is co-spatial with 11.3A while the 16.4~\mum band is not.
For the 12.7~\mum band, considering the blue 12.7A sub-component individually has a major effect on the relationship with the 17.4~\mum band.
The 12.7A feature relates to 17.4~\mum analogously to 11.3A, despite the fact that the total 12.7~\mum emission relates best to the 16.4~\mum band.

These somewhat subtle differences in the correlation diagram originate from differences which are more obvious when the spatial maps (Fig.~\ref{fig:ratiomaps}) or the cuts (Fig.~\ref{fig:cuts}) are inspected visually, emphasizing the importance of these spatially resolved studies.
Summarizing, the diagrams of Fig.~\ref{fig:corr} revealed a key difference between the 16.4 and 17.4~\mum bands: 17.4~\mum emission is co-spatial with the 7.7 narrow, 8.6, and 11.0 features and the blue 11.3 and 12.7~\mum sub-components, while only part of the contributions to
the 16.4~\mum band come from the same population.
In the following sections we discuss how these connections could be interpreted.

\subsection{Charge tracers}
\label{sec:charge}

What several of the features highlighted in the previous section have in common, is that they are considered tracers of charged PAHs.
The 6.2/11.3, 7.7/11.3, 8.6/11.3, and 11.0/11.3 ratios are considered key quantities to trace the ionization state of PAHs, where it is assumed that the 11.3~\mum emission is dominated by emission from neutrals, while the 6.2, 7.7, 8.6, and 11.0 bands are dominated by cationic PAHs \citep{Hudgins1999, Hony2001}.
These tracers were used previously to interpret the changing band ratios across NGC\,7023 in terms of the charge state \citep{Shannon2015, Stock2016, Boersma2014, Peeters2017, Sidhu2022}.
In Fig.~\ref{fig:corr}, the region including ATM up to DF1 appears in the top right of the correlation diagrams, indicating that charged emission carriers are more abundant, as expected from previous work.
Our maps (Figs.~\ref{fig:mapstotals}-\ref{fig:maps127}) and spatial profiles (Fig.~\ref{fig:cuts}) paint a spatially resolved picture about the contribution of charged and neutral carriers to the different bands.
We suggest that the type III features are dominated by a charged population that itself has a type III spatial distribution, while the neutral population is distributed according to type I.
The mixed (type II) features then contain contributions by both charged and neutral carriers in varying ratios.

Previously, \citet{Boersma2014} revealed that the spectra can be grouped into morphological zones that run parallel to the NGC\,7023 DF, and concluded that both cationic and neutral PAHs contribute significantly to the 7.7 and 12.7~\mum bands.
Other works include blind signal separation techniques, which revealed three representative spectra and spatial distributions in NGC\,7023 \citep{Berne2007, Rosenberg2011}, which the authors associated with charged PAHs, neutral PAHs, and VSGs.
\citet{Boersma2015} discuss the use of the ``traditional'' charge tracers, and concluded that the 7.7/11.3 ratio is not a reliable charge tracer near the DF.
Our work shows that using the 7.7~\mum band as a charge tracer requires carefully separating the narrower (type III) sub-components from the broad (type I) 7.7~\mum emission.
Studies of the effectiveness of certain band ratios as charge tracers, such as the one by \citet{Boersma2015}, could be revisited while taking the individual sub-components under consideration.
An in-depth comparison of charge tracers may also reveal the selective destruction of certain sub-populations within the cation population \citep{Singh2025}.

Via the correlation with 11.0~\mum, \citet{Shannon2015} attributed the 17.4~\mum feature to cationic PAHs, and suggested that the 16.4 band arises from both cationic and neutral PAHs.
Our results support that the 16.4~\mum band is of a more mixed nature (type II) than the 17.4~\mum feature (type III), and we showed that the latter was more strongly correlated to the ionized PAH tracers (Fig.~\ref{fig:corr}).
\citet{Shannon2016} studied the 11.3 and 12.7~\mum profile variations in NGC\,7023 and NGC\,2023, and report a cationic contribution to 11.3~\mum.
Similarly, they found a potentially neutral contribution to the 11.0~\mum band by decomposing it into two components.
Their two components for the 12.7~\mum band have different spatial distributions as well, and one of them is a sharper peak on the blue side.
Our JWST-based decomposition and maps now resolve these spatial distribution differences: the 12.7A and 12.7B components have type III and type I spatial distributions respectively, with most of the profile variation taking place over a distance of less than $1\arcsec$ as revealed by the 12.7A/12.7 ratio map (Fig.~\ref{fig:ratiomaps}).
A similar effect is found for the 11.3~\mum band: while the total 11.3~\mum emission is often used to trace neutral PAHs, our work reveals that the 11.3A component contributes up to 20\% to this band.

\subsection{Profile diagnostics and comparison to the Orion Bar}
\label{sec:compareorion}

The PDRs4All program \citep[ERS-1288]{Berne2022}  observed the Orion Bar PDR with a similar observing strategy as our NGC\,7023 observations.
Several previous works analyzing the PDRs4All spectra made use of the ``template spectra'' that were extracted from five apertures representing the key zones in this PDR \citep[available via the CDS]{Peeters2024, Chown2024, VanDePutte2025}.
They include the ionized region ahead of the IF (named ``H{\sc{ii}}'', but it also has a PDR in the background), the neutral region between the IF and the DF (``atomic PDR''), and three DF (``DF1'', ``DF2'', ``DF3'').
\citet{VanDePutte2025} applied the PDR pack to these five spectra and discussed how PAHFIT captures a number of band profile variations which can serve as diagnostics for the photochemical evolution of the emission carriers.
One of the results includes a diagram comparing the 6.2~\mum FWHM to the 7.7 broad/7.7 total ratios, showing that both of these increase for the more shielded DF1, DF2, and DF3 regions.
This was interpreted as follows: the 6.2~\mum profile, associated with pure aromatic C-C stretching modes and enhanced for cationic species \citep{Allamandola1999, Peeters2002}, broadens due to the effects of anharmonicity \citep{Mackie2022} when small PAHs with higher excitation temperatures contribute to the emission.
The 7.7 broad contribution to the total 7.7~\mum emission is thought to originate from a population of VSGs \citep{Berne2007, Pilleri2012, Peeters2017, Stock2017, Khan2025}.
Both of these components are expected to be vulnerable to the UV radiation field, and they are therefore enhanced in the shielded regions.

\begin{figure}[tb]
    \centering
    \includegraphics[scale=1.1]{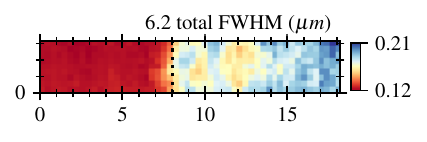}
    \includegraphics{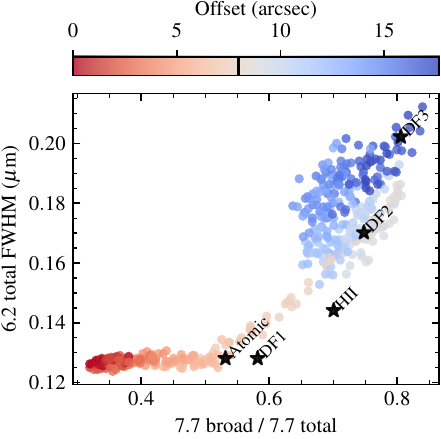}
    \caption{Profile diagnostics diagram based on the 6.2 and 7.7~\mum profiles, comparing our results for NGC\,7023 (colorful circles) with those for the five Orion Bar template spectra  \citep[labeled markers,][]{VanDePutte2025}.
    The color bar indicates the position in the NGC\,7023 mosaic and the black marker bar is the location of DF1.
    In the atomic PDR regime, the NGC\,7023 spectra exhibit much more strongly pronounced 7.7~\mum narrow features compared to the Orion Bar ``Atomic'' template.
    The 6.2~\mum FWHM tends to a constant value, and the upper panel shows a map of the 6.2~\mum FWHM with DF1 marked by the dotted line.
    For the more shielded regime (NGC\,7023 beyond DF1; Orion Bar HII, DF2, and DF3), the data points follow a similar path in the diagram for both objects.
    }
    \label{fig:diag6277}
\end{figure}

We reproduced this diagram in Fig.~\ref{fig:diag6277} and applied the same diagnostics to the NGC\,7023 spectra, to clarify how the observed profiles and conditions compare to each of the Orion Bar template regions.
The 6.2~\mum FWHMs were calculated analogously to \citet{VanDePutte2025}, by evaluating the 6.2~\mum profile of the PAHFIT model (sum of Drude profiles, Table.~\ref{tab:legend}), and determining the maximum and half-maximum points numerically.
The Orion Bar templates are ordered in the following way for all diagnostics examined by \citet{VanDePutte2025}: atomic PDR, DF1, H{\sc{ii}} (background PDR), DF2, DF3.
The point cloud for NGC\,7023 in Fig.~\ref{fig:diag6277} is color-coded by the distance from the edge of the mosaic, and a similar sequence is found, where DF1 and the region beyond show enhanced 6.2 FWHM and 7.7 broad/7.7 total values.
The DF1 spectra of NGC\,7023 (gray dots as marked with the black line on the color bar) appear to be most similar to the conditions of DF2 in the Orion Bar.
The narrowing of the 6.2~\mum band toward the star tends to the same FWHM value as in the Orion Bar.

In our decomposition, the 7.7~\mum broad feature has a strikingly similar spatial distribution to that of the red wing of the 11.3~\mum band (11.3D, Fig.~\ref{fig:maps112}).
Three spectral signals were extracted over the 10.0-19.5 \mum range from NGC\,7023 Spitzer spectra by \citet{Rosenberg2011}, and they were characterized using PAHdb.
They showed a first component with a pronounced 11.0~\mum band, a second component with a pronounced but narrower 11.3~\mum band, and a third component with a broader red wing at 11.3~\mum, interpreted as PAH cations, neutral PAHs, and VSGs respectively.
For the Orion Bar, \citet{Khan2025} decomposed the 11.3~\mum band into two profiles extracted from the data, both of which have an asymmetric nature, and find that the red wing of the 11.3~\mum band behaves more independently from the main PAH emission bands, suggesting VSGs or PAH clusters as a possible origin.
\citet{Boersma2014} include the following candidates for the red wing of 11.3~\mum:
PAH anions, anharmonicity of hot PAHs, VSGs and PAH clusters, or potentially superhydrogenation near DF1, which can play a role in the H$_2$ formation taking place \citep{Rauls2008, Thrower2012, Jensen2019}.
Another suggestion for the shift of the 7.7~\mum band (here facilitated by the broad contribution) is the presence of aliphatic side groups, which are expected to be removed when the UV radiation field increases \citep{Bregman2005, Sloan2007, Sadjadi2015a, Shannon2019}.
That could explain the similarity of 7.7 broad with the 3.4~\mum profile (aliphatic C-H stretch, Fig.~\ref{fig:cuts}).

The regions close to the star exhibit much lower values of 7.7 broad / 7.7 total; even when compared to the Orion Bar atomic PDR (Fig.~\ref{fig:diag6277}).
The narrow features continue getting more pronounced as the star is approached, while the broad component is further suppressed.
The ratios of 11.3A / 11.3, 12.7A / 12.7, and 17.4 / 16.4 in Fig.~\ref{fig:ratiomaps} exhibit a similar gradient, and this cohesive behavior between the band profiles suggests that there are at least two populations, one which becomes gradually more dominant closer to the star.
We also note that the 11.0~\mum band in the NGC\,7023 ATM region (Fig.~\ref{fig:maps112}) is stronger relative to 11.3~\mum, compared to what is seen in the Orion Bar \citep{Khan2025}.
All of the above shows that the differences in the emitting populations are more pronounced than what is observed in the Orion Bar, and we speculate about the cause as follows.

The harder radiation field driving the Orion Bar PDR results in the formation of an IF, which is not the case for the softer radiation field in NGC\,7023. 
This may result in different behavior of the PAH size and charge as a function of the radiation field, for which the intensity is determined by the distance to the star or the shielding by the gas and dust when entering the PDR.
The harder radiation field will dehydrogenate and destroy PAHs and ionize the surviving population, which may lead to a situation where only larger species survive in the region with a significant ionized PAH fraction.
The PAH database fitting approach by \citet{Maragkoudakis2026} finds that only large PAHs ($> 70$ C atoms) survive near the IF of the Orion Bar.
While the estimated PAH size in NGC\,7023 also increases closer to the star \citep[e.g.,][]{Croiset2016}, the softer radiation field may allow a substantial population of smaller PAH to survive in the region where PAHs are ionized, and such smaller PAH are dehydrogenated more quickly \citep{Andrews2015}. 
The presence of these smaller PAHs may lead to a different response of the charged/neutral PAH tracers (such as the narrow 7.7~\mum features or larger 11.0/11.3 ratio of NGC\,7023). 

Besides the radiation field, the hydrogen density will also affect the dehydrogenation balance, as higher densities restore the hydrogenation at a faster rate.
The different regions in NGC\,7023 exhibit a large range of densities, from the order of $10^2$ cm$^{-3}$ in the central cavity, to $10^4$ cm$^{-3}$ near the PDR surface \citep{Pilleri2012}, up to $10^5$ or $10^6$ cm$^{-3}$ for local clumps and filaments \citep{Koehler2014}.
Chemical modeling of NGC\,7023 by \citet{Murga2022} showed that including the dehydrogenation balance changes the PAH evolution depending on the assumed age of the nebula: a major fraction of the PAHs survive after $10^4$ years, while most are destroyed by $10^5$ years.
Detailed photochemical modeling \citep{Montillaud2013, Andrews2015, Murga2022}, consistent with the constraints provided by our data, will be necessary to understand the interplay between the processes and how the environmental conditions lead to the observed outcome.

\subsection{Outlook: the path to fullerenes}
\label{sec:outlook}

The central region of NGC\,7023 is known to exhibit fullerene (C$_{60}$) emission within a distance of $\sim$25\arcsec\ from HD200775 \citep{Sellgren2010, Berne2012}.
This detection with Spitzer was based on the main features of C$_{60}$ at 7.05, 7.5, 17.4, and 18.9~\mum \citep{Cami2010}.
None of these characteristic features are observed in the spectra of this GTO program, confirming that no C$_{60}$ is formed this close to the DF.
While we do observe a 17.4~\mum feature that gets stronger toward the star, this feature can also be explained by PAHs that emit at this wavelength \citep{Sellgren2010, Shannon2016}.
In this work we emphasize the difference between the spatial distributions of the 17.4~\mum and 16.4~\mum features, and how the 17.4~\mum band becomes gradually more prominent while the profiles of several other bands evolve alongside it (5.7, 6.2, 7.7, 11.3, 12.7~\mum).
The band ratio gradients shown in Fig.~\ref{fig:ratiomaps} suggest that the changes in the band profiles will continue beyond the edge of our spectroscopic field of view.
This suggests that observing regions progressively closer to the star would trace  further changes in the carbonaceous emission bands alongside the appearance of the \csixty emission, which would allow a characterization of the photoprocessing stages that eventually lead to the formation of fullerenes.
Below, we briefly summarize a previously suggested fullerene formation pathway, where PAHs are processed into fullerenes \citep{Berne2012, Andrews2015}.
Then we provide an outlook with some suggestions as to which spectral details could contain information about the processing stages taking place.

On the irradiated side of DF1, the ``grandPAH'' scenario may be taking place \citep{Andrews2015}, in which PAHs are created when the UV radiation field evaporates them from other existing materials such as VSGs or PAH clusters \citep{Rapacioli2005, Pilleri2012}.
The initial PAH species are processed into a population of ``grandPAHs'', a smaller set of species consisting of more photostable PAHs which are larger, more compact, and more symmetric.
This scenario was proposed to explain how the carbonaceous band spectrum appears mostly insensitive to the radiation field parameter ($G_0$) driving different PDRs \citep{Andrews2015}.

The fullerene formation scenario of \citet{Berne2012} involves further photoprocessing of the initial population of larger PAHs.
For the conditions in the area where C$_{60}$ is observed, the PAH photoprocessing models by \citet{Montillaud2013} suggest that PAHs with a typical size of 50-70 carbon atoms can be fully dehydrogenated through photodissociation of the H atoms, leaving a population of similarly sized carbon clusters.
The next photoprocessing steps for the formation of fullerenes then involve the removal of carbon atoms from the hexagonal grid, which introduces five-membered rings that allow the fragments to curve \citep{Berne2012, Berne2015}.
When such fragments form closed cages, lab experiments confirm that they can shrink via the removal of C-C pairs, until a stable fullerene is formed \citep{Zhen2014}.

An alternative fullerene formation scenario involves the processing of hydrogenated amorphous carbon grains (HAC), observed as certain spectral plateaus at 6-9 and 10-13~\mum in planetary nebulae that contain fullerenes \citep{BernardSalas2012}.
Photoprocessing the larger HAC grains creates structures of a mixed aromatic and aliphatic nature, and five-membered rings are formed by dehydrogenating some of those structures \citep{GarciaHernandez2010, Micelotta2012}.
Such species have been named ``proto-fullerenes'', and may emit spectral lines that are normally associated with C$_{60}$.
The 16.4~\mum band is also considered an indicator of five-membered rings \citep{Duley2012}.

While the 16.4 and 17.4~\mum bands are both associated with skeletal C-C-C modes of PAHs, their spatial distributions are different (Sect.~\ref{sec:discuss164}).
One possible interpretation was via their correlation with typical PAH charge tracers (Fig.~\ref{fig:corr}).
One could expect that there must be certain structural differences between the neutral and charged population, so that the charged one dominates the 17.4~\mum emission.
Nonetheless, \citet{Boersma2010} reported that the PAH charge is not expected to strongly affect the 16-18~\mum spectra.
Their search within their spectral database for species with strong bands near 16.4 or 17.4~\mum did not reveal a structural class that preferentially emits at 17.4~\mum.
Combining recent expansions of the database \citep{Bauschlicher2018, Mattioda2020} with the depth and spatial resolution of JWST at 16-18~\mum may lead to different conclusions.

Additional information about the processing may be available from the AIBs in the 10-15~\mum range associated with C-H out-of-plane bending modes of PAHs.
These bands are assigned to vibrational modes involving the different hydrogen adjacency classes (solo, duo, trio, quartet, quintet), which trace the edge structure of PAHs \citep{Hudgins1999, Hony2001}.
The ratios of these modes can be modified by changes in the edge structures of the carbon skeleton or by dehydrogenation, potentially a first step in the fullerene formation process \citep{Berne2012}.
\citet{Fleming2010} produced maps of the same region using the PAHFIT IDL routine, and interpreted the 12.7/11.3 ratio and the associated hydrogen adjacency as an indicator of the dehydrogenation state.
They found that this ratio is higher both at the bright filament and close to the star, and reaches a minimum about 10\arcsec~away from the DF in the star direction.
However, other work shows that even though 11.3~\mum band is dominated by solo modes, and 12.7~\mum by duo or trio modes \citep[e.g.,][]{Khan2025}, the 12.7/11.3 variation in NGC\,7023 is dominated by charge effects \citep{Rosenberg2011, Peeters2012, Boersma2015, Shannon2016}.
The blue side of the 11.3~\mum band seems to be related to charge tracers (Sect.~\ref{sec:charge}), but other interpretations are possible, such as the temperature or mass distribution \citep{Candian2015}, or the presence of molecules containing oxygen or magnesium  \citep{Sadjadi2015}.

Changes in the PAH size are also a piece of the puzzle, as the carbon fragments in the scenario by \citet{Berne2012} need to be larger than 60 atoms.
A variety of size tracers will be explored in upcoming work (E. Roscoe et al., in preparation).
Previous PAH size investigations in NGC\,7023 suggest that small PAHs are broken down in the cavity and only large species survive \citep{Boersma2013, Boersma2014, Croiset2016}.
Our map of the 17.4~\mum band would support this, as previous work based on PAHdb found that compact PAHs of 50-130 carbon atoms consistently show a feature near this wavelength \citep{Boersma2010}.
In Sect.~\ref{sec:compareorion} an initial look into the size was provided through the 6.2~\mum FWHM map of Fig.~\ref{fig:diag6277}, where the narrower 6.2~\mum band was interpreted as larger PAH sizes.
It appears that the 6.2~\mum FWHM tends to a constant minimum in the direction of the star, though this may also indicate that this tracer becomes insensitive to the PAH size at that point.

A JWST program that completed during Cycle 4 (ID: GO-8007; PI: Tielens) has observed three regions in the cavity of NGC\,7023 with MIRI MRS, with a primary goal of characterizing the C$_{60}$ cations.
These spectra will also reveal how much further the AIB profiles keep evolving, further probing the processing sequence of the AIB carriers between the NGC\,7023 NW filament and the star.

\section{Summary and conclusions}

We apply the PAHFIT spectral decomposition tool to extract the brightness of individual sub-components of the carbonaceous emission bands observed with the NIRSpec and MIRI IFUs in NGC\,7023.
The spatially resolved maps of the individual features presented in this work reveal that the emission band sub-components have distinct spatial distributions.
We organized the features into types I, II, and III, according to their spatial distributions, where the roman numeral is an indicator of the brightness in the atomic PDR region (ATM) relative to the main dissociation front (DF1), in ascending order (Sect.~\ref{sec:types}, Table~\ref{tab:types}).

The 17.4~\mum band in particular is of type III, which contrasts it to the type I or II nature of most other emission bands, if only the sum of all their sub-components is considered.
However, mapping the individual sub-components of the 5.7, 7.7, 11.3, and 12.7~\mum bands reveals that their sub-components have different spatial distributions.
Consistently, the blue-side sub-components such as 11.3A are of type III, even when the redder sub-components or the total emission of each band is of type I or II.
The spatial distributions indicate the existence of at least two particular populations of carbonaceous emission band carriers, one of which is strongly emitting in the ATM region, while the other is suppressed relative to DF1.
Normalizing the bands to the total 11.3~\mum emission, reveals strong correlations of the 17.4~\mum band with these sub-components, and with the 8.6 and 11.0~\mum features as well, which are considered tracers of ionized PAHs. This indicates a connection between the 17.4~\mum band and a population of ionized carbonaceous emitters.

The changes in the populations may relate to the photochemical evolution of PAHs, which is driven to a more advanced stage by the increasing radiation field as the star is approached.
Maps of a set of emission band profile indicators (5.7A/5.7, 7.7 narrow/7.7 broad, 11.3A/11.3, 12.7A/12.7; Fig.~\ref{fig:ratiomaps}) all exhibit a positive gradient in the direction of the star that persists up to the edge of the field of view, indicating that the profiles will keep evolving for smaller distances.
Since \csixty emission is known to be present closer to the star, these profile changes may indicate processing steps that are part of a fullerene formation pathway.
Recently completed JWST spectroscopy along this spatial direction will reveal the continuation of the evolution sequence revealed in this work.

As JWST observes more spectra of similar quality, the variations in the emission profiles and their relationship to the environmental conditions will be strongly constrained.
It is therefore increasingly important that theoretical models can explain profile variations or the spatial distribution of individual sub-components --rather than ratios of total band intensities--
and the interpretation in terms of the underlying populations should be consistent with the environment (radiation field, density, presence of atomic or molecular hydrogen).
In particular, determining which photochemical processes near DF1 cause the changes in the 17.4/16.4 ratio will require models that accurately predict the emission features in this wavelength range.

\section{Data availability\label{sec:dataavail}}
Machine-readable versions of the maps presented in Figs.~\ref{fig:mapstotals}, \ref{fig:maps57}, \ref{fig:maps77}, \ref{fig:maps112}, \ref{fig:maps127}, and Fig.~\ref{fig:allmaps} are only available in electronic form at the CDS via \url{http://cdsarc.cds.unistra.fr/viz-bin/cat/J/A+A/xxx/yyy}.

\begin{acknowledgements}
This work is based on observations made with the NASA/ESA/CSA James Webb Space Telescope. 
The data were obtained from the Mikulski Archive for 
Space Telescopes at the Space Telescope 
Science Institute, which is operated by the Association of Universities for Research in 
Astronomy, Inc., under NASA contract NAS 5-03127 for JWST. These observations are associated 
with GTO program \#01192.
D. VDP, K.D.G, and A.N-C are partially supported by NASA grant 80NSSC21K1294.
K.M. is supported by JWST–NIRCam contract no. NAS5-02015 to the University of Arizona.
MB acknowledges funding from the Belgian Science Policy Office (BELSPO) through the PRODEX 
project “JWST/MIRI Science exploitation” (C4000142239).
Part of this work was performed at the French MIRI center with the support of 
CNES and the ANR-labcom INCLASS between IAS and ACRI-ST, and also supported 
by the Programme National ``Physique et Chimie du Milieu Interstellaire'' (PCMI) of 
CNRS/INSU with INC/INP co-funded by CEA and CNES.
This work made use of Astropy:\footnote{\url{http://www.astropy.org}} a community-developed core
Python package and an
ecosystem of tools and resources for astronomy \citep{AstropyCollaboration2013,AstropyCollaboration2018,AstropyCollaboration2022}.

\end{acknowledgements}

\bibliographystyle{aa}
\bibliography{7023paper}

\begin{appendix}

\section{Overview of all maps}
\label{app:allmaps}

\begin{figure}[hbp]
    \centering
    \includegraphics[]{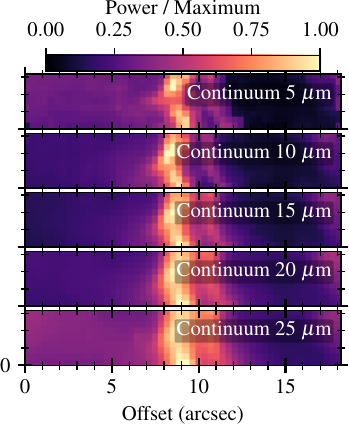}
    \caption{Maps of the continuum model evaluated at different wavelengths (as indicated in by the labels in \mum).
    Each map is normalized to its maximum value.}
    \label{fig:contmaps}
\end{figure}

\begin{figure*}[tbp]
    \includegraphics[width=0.24\linewidth]{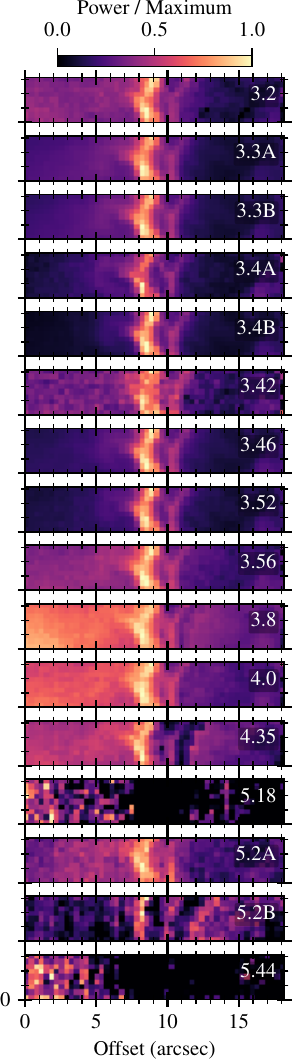}
    \includegraphics[width=0.24\linewidth]{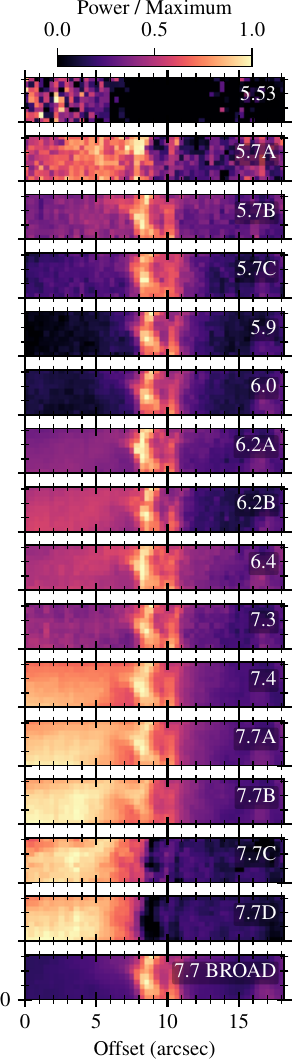}
    \includegraphics[width=0.24\linewidth]{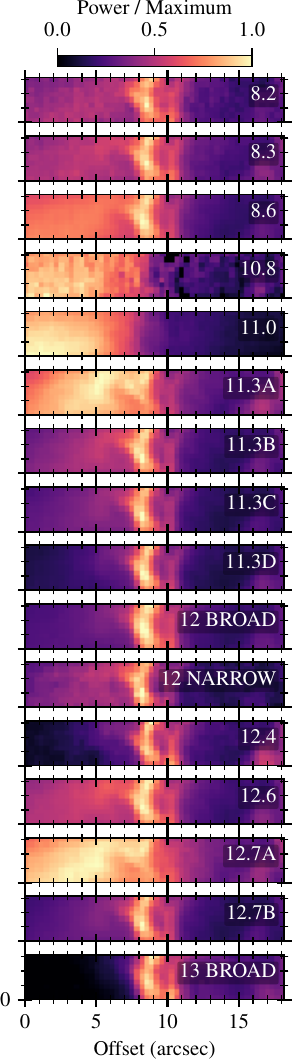}
    \includegraphics[width=0.24\linewidth]{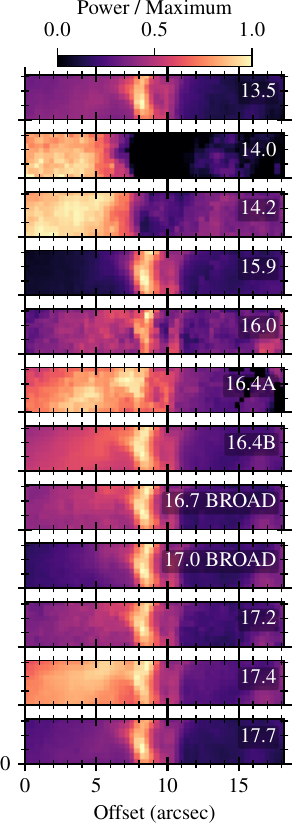}
    \caption{Full overview of spatial distributions extracted from PAHFIT results.
    The maps show normalized quantities analogous to Fig.~\ref{fig:mapstotals}: the frequency-integrated surface brightness (W m$^{-2}$ sr$^{-1}$), divided by the maximum value of the map.}
    \label{fig:allmaps}
\end{figure*}

For completeness, in Fig.~\ref{fig:contmaps}, we illustrate the spatial distribution of the dust continuum emission as derived from the PAHFIT results.
The results themselves consist of a set of best-fit scaling factors for the MBB component at each temperature.
The continuum maps shown in Fig.~\ref{fig:contmaps} were created by taking the sum of the continuum model functions, and evaluating the resulting continuum model at the given wavelengths.
Additional feature intensity maps are shown in Fig.~\ref{fig:allmaps}.
The wavelength and FWHM of each feature are given in the science pack published by \citet{VanDePutte2025}, and modified by the updates in Table~\ref{tab:tune16}.
A machine readable version of these maps is available via the CDS (Sect.~\ref{sec:dataavail}).

\section{Fit residuals}

\begin{figure*}[tp]
    \centering
    \includegraphics[scale=0.9]{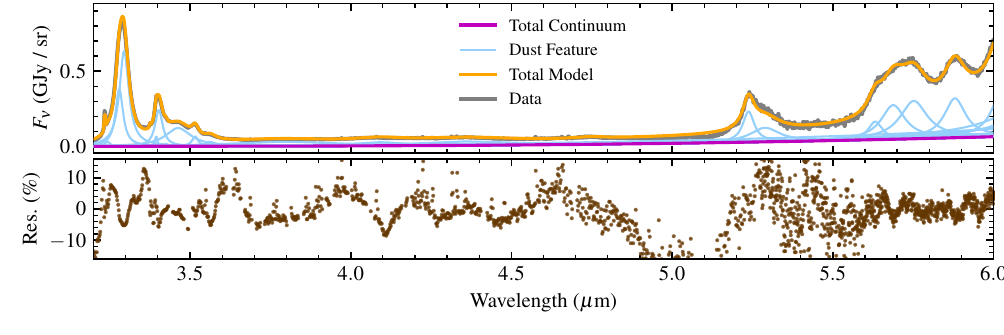}
    \includegraphics[scale=0.9]{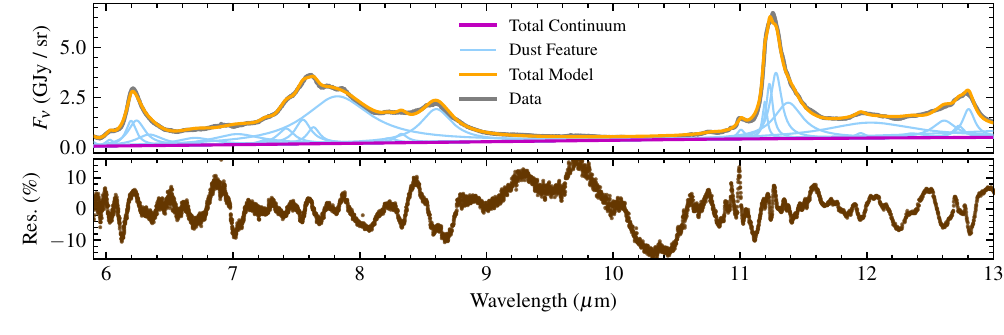}
    \includegraphics[scale=0.9]{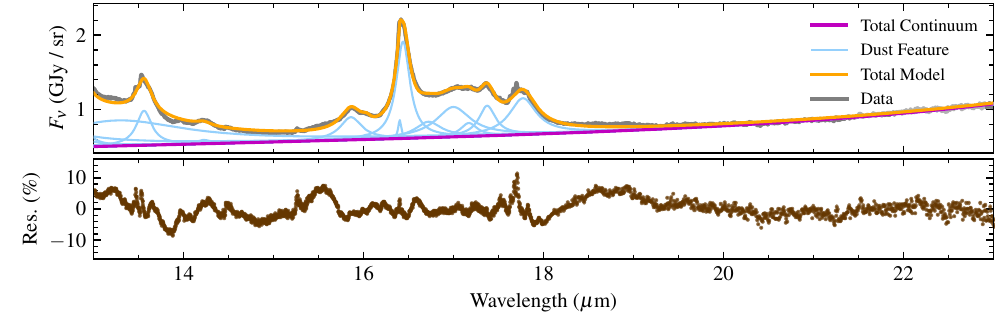}
    \caption{Residuals of the DF1 fit (same as in Figs.~\ref{fig:specandfit} and \ref{fig:specandfit2}), defined as (data - model) / model.
    The largest deviations are in the continuum near the wings of the main emission complexes.}
    \label{fig:residuals}
\end{figure*}

In Fig.~\ref{fig:residuals}, we show the residuals of the DF1 fit of Figs.~\ref{fig:specandfit} and \ref{fig:specandfit2}.
The model approximates the spectrum within 10\%, with larger deviations up to 15\% taking place in the wavelength ranges of 4.5-5.2~\mum and 9-11~\mum.
Besides fitting performance, these figures also indicate the potential presence of additional profile variations or sub-components, such as the peaks near 6.9 and 9.8~\mum.

\end{appendix}

\end{document}